\documentclass[10pt]{article}
\usepackage[USenglish]{babel}
\usepackage{a4wide}
\usepackage{epsfig,color}
\usepackage[all]{xy}
\usepackage{amssymb}
\usepackage{amsmath}
\usepackage{latexsym}
\usepackage{pifont}
\usepackage{pdfsync}
\usepackage{enumerate}
\usepackage{setspace}
\usepackage{euler}
\usepackage[latin1]{inputenc}
\title{Using Coalgebras and the Giry Monad for Interpreting Game Logics --- A Tutorial}
\author{Ernst-Erich Doberkat\footnote{Some of the results were obtained while the author held the Chair for Software Technology at Technische Universität Dortmund; they were funded in part by Deutsche Forschungsgemeinschaft, \emph{Koalgebraische Eigenschaften stochastischer Relationen}, grant DO 263/12-1.}\\\textsc{Math ++ Software}, Bochum, Germany\\\texttt{math@doberkat.de}}
\date{\today}
\newcommand{\labelImpl}[2]{\ensuremath{\ref{#1}~\Rightarrow~\ref{#2}}}

\newcommand{\Klasse}[2]{\left[#1\right]_{#2}}
\newcommand{\Faktor}[2]{{#1}/{#2}}

\newcommand{\Bild}[2]{{#1}\left[#2\right]}
\newcommand{\InvBild}[2]{\Bild{#1^{-1}}{#2}}
\newcommand{\Kern}[1]{\mathsf{ker}\left(#1\right)}
\newcommand{\Folge}[1]{(#1_n)_{n \in \Nat}}

\newcommand{\theTheory}[2]{Th_{#1}({#2})}
\newcommand{\spaceFont}[1]{\mathfrak{#1}}

\newcommand{\Prob}[1]{\spaceFont{P}\left(#1\right)}
\newcommand{\SubProb}[1]{\spaceFont{S}\left(#1\right)}

\newcommand{\Category}[1]{\ensuremath{\mathbf{#1}}}
\newcommand{\SubProbSenza}{\spaceFont{S}}
\newcommand{\ProbSenza}{\spaceFont{P}}
\newcommand{\PowerSet}[1]{\ensuremath{\mathcal{P}\left(#1\right)}}
\newcommand{\PowerSenza}{\ensuremath{\mathcal{P}}}
\newcommand{\FunctorSenza}[1]{\ensuremath{\spaceFont{#1}}}

\newcommand{\Borel}[1]{\ensuremath{{\mathcal B}(#1)}}

\edef\LinkeKlammer{\lbrack\!\lbrack}
\edef\RechteKlammer{\rbrack\!\rbrack}
\newcommand{\Gilt}[1][\phi]{\ensuremath{\LinkeKlammer#1\RechteKlammer}}

\newcommand{\Trans}{\rightsquigarrow}

\newtheorem{definition}{Definition}[section]
\newcommand{\BeginDefinition}[1]{%
  \begin{definition}\label{#1}
}
\newcommand{\EndDefinition}{\end{definition}}

\newtheorem{example}[definition]{Example}
\newcommand{\BeginExample}[1]{%
  \begin{example}\label{#1}\rm
}
\newcommand{\EndExample}{--- \end{example}}

\newtheorem{observation}[definition]{Observation}
\newcommand{\BeginObservation}[1]{
  \begin{observation}\label{#1}\rm
}
\newcommand{\EndObservation}{--- \end{observation}}

\newtheorem{theorem}[definition]{Theorem}
\newcommand{\BeginTheorem}[1]{%
  \begin{theorem}\label{#1}
}
\newcommand{\EndTheorem}{\end{theorem}}

\newtheorem{corollary}[definition]{Corollary}
\newcommand{\BeginCorollary}[1]{
  \begin{corollary}\label{#1}
}

\newtheorem{proposition}[definition]{Proposition}
\newcommand{\BeginProposition}[1]{%
  \begin{proposition}\label{#1}
}
\newcommand{\EndProposition}{\end{proposition}}

\newcommand{\EndCorollary}{\end{corollary}}
\newtheorem{lemma}[definition]{Lemma}
\newcommand{\BeginLemma}[1]{%
  \begin{lemma}\label{#1}
}
\newcommand{\EndLemma}{\end{lemma}}

\newtheorem{claim}{Claim}
\newcommand{\BeginClaim}[1]{%
  \begin{claim}\label{#1}
}
\newcommand{\EndClaim}{\end{claim}}

\newenvironment{proof}{\textbf{Proof\ }}{\ensuremath{\QED}}
\newcommand{\BeginProof}{\begin{proof}}
\newcommand{\EndProof}{\end{proof}}

\newenvironment{remark}{\textbf{Remark:\ }}{}
\newcommand{\BeginRemark}{\begin{remark}}
\newcommand{\EndRemark}{\QED\end{remark}}
\newcommand{\QED}{%
\ensuremath{\dashv}
}

\newcommand{\Real}{\mathbb{R}}

\newcommand{\Nat}{\mathbb{N}}
\newcommand{\Rational}{\mathbb{Q}}

\def\mathsf{\mathrm}
\def\theta{\vartheta}
\def\dots{\ldots}

\makeatletter
\def\@axx#1{\ensuremath{\mathbb{(#1)}}}
\def\AC{\@axx{AC}}
\def\WO{\@axx{WO}}
\def\ZL{\@axx{ZL}}
\def\MP{\@axx{MP}}
\def\MI{\@axx{MI}}
\def\AD{\@axx{AD}}
\makeatother

\newcommand{\isEquiv}[3]{\ensuremath{{#1}\ {#3}\ {#2}}}
\newenvironment{theExercises}
{\begin{exercise}\rm}
{\end{exercise}}
\newtheorem{exercise}{Exercise}
\newcommand{\BeginExercise}[1]{%
  \begin{theExercises}\label{#1}
}
\newcommand{\EndExercise}{\end{theExercises}}

\def\endEx{{\Large\ding{44}}}
\renewcommand{\BeginExample}[2][\empty]{%
  \begin{example}\label{#2}\rm
\ifx#1\empty\empty \else\write\beispielIndex{\noexpand\bspEintrag(#2)({#1})}\fi
}
\renewcommand{\EndExample}{\endEx \end{example}}
\def\exComment (#1)(#2){#1 (p.~\pageref{#2})}

\newcommand{\BeginExxample}[2][\empty]{%
  \begin{example}\label{#2}$\circledS$\rm
\ifx#1\empty\empty \else\write\beispielIndex{\noexpand\bspEintrag(#2)(#1)}\fi
}
\newcommand{\EndExxample}{\endEx \end{example}}

\def\CatFont{\mathbf}
\def\Category#1{\ensuremath{\CatFont{#1}}}
\renewcommand{\spaceFont}[1]{\CatFont{#1}}

\def\catC{\Category{C}}

\def\catSET{\Category{Set}}

\def\funF{\FunctorSenza{F}}

\def\funT{\FunctorSenza{T}}
\def\funV{\FunctorSenza{V}}

\makeatletter
\def\@theProbSenza#1{\ensuremath{\mathbb{#1}}}
\def\SubProbSenza{\@theProbSenza{S}}
\def\ProbSenza{\@theProbSenza{P}}
\def\FinSenza{\@theProbSenza{M}}
\def\SigmaSenza{\@theProbSenza{M}_{\ensuremath{\sigma}}}
\makeatother
\renewcommand{\Prob}[1]{\ProbSenza\left(#1\right)}
\def\SubProb#1{\SubProbSenza(#1)}
\def\FinM#1{\FinSenza(#1)}

\makeatletter
\def\@beta{\beta}
\def\betaSenza#1{\@beta_{#1}}
\def\theBeta#1#2#3{\ensuremath{\betaSenza{#3}(#1, #2)}}

\makeatother

\newcommand{\BorelSenza}{\ensuremath{{\mathcal B}}}
\renewcommand{\Borel}[1]{\ensuremath{\BorelSenza(#1)}}

\def\MeasbFnct#1{\ensuremath{{\cal F}(#1)}}

\def\Graph#1{\ensuremath{\mathsf{graph}(#1)}}
\def\DefSect{
\ifthenelse{\boolean{isBook}}{
\def\Section{\chapter}
\def\Subsection{\section}
\def\Subsubsection{\subsection}
\def\Subsubsubsection{\subsubsection}
}{
\def\Section{\section}
\def\Subsection{\subsection}
\def\Subsubsection{\subsubsection}
\def\Subsubsubsection{\paragraph}
}
}

\makeatletter
\def\@Tut#1#2{\begin{color}{red}12345\cite[#1]{#2}\marginpar[$\dots\Longrightarrow\dots$]{$\dots\Longleftarrow\dots$}\end{color}}
\def\CategCite#1{\@Tut{#1}{EED-Categs}}
\def\SetCite#1{\@Tut{#1}{EED-Tut_sets}}
\def\TopCite#1{\@Tut{#1}{EED-Topologies}}

\makeatother

\makeatletter
\def\@norm#1#2#3{\ensuremath{||{#1}||_{#2}^{#3}}}
\def\infNorm#1#2{\@norm{#1}{\infty}{#2}}
\def\aNorm#1{\@norm{#1}{}{}}
\makeatother

\makeatletter
\def\@conv#1{\stackrel{#1}{\longrightarrow}} 
\def\aeC{\@conv{a.e.}} 
\def\nmC{\@conv{i.m.}}
\makeatother

\makeatletter
\newcommand{\@theL}[3]{\ensuremath{{#3}_{#1}{#2}}}
\newcommand{\cLp}[2][p]{\@theL{#1}{(#2)}{{\cal L}}}
\newcommand{\rLp}[2][p]{\@theL{#1}{(#2)}{L}}
\newcommand\cLpS[1][p]{\@theL{#1}{}{{\cal L}}}
\newcommand\rLpS[1][p]{\@theL{#1}{}{L}}
\makeatother

\renewcommand{\Bild}[2]{{#1}\bigl[#2\bigr]}
\renewcommand{\Folge}[2][n]{\ensuremath{({#2}_{#1})_{#1\in\Nat}}}

\makeatletter
\newcommand{\@imText}[2][b]{%
\marginpar[{\parbox[#1]{\marginparwidth}{#2}}]{\parbox[#1]{\marginparwidth}{\raggedleft{#2}}}%
}
\definecolor{@leichtgrau}{gray}{.60}
\def\Rot#1{\begin{color}{red}#1\end{color}}
\newcommand{\@randMarkierung}[1]{
\mbox{}\marginpar{\raggedleft\hspace{0pt}\textsf{\Rot{#1}}}}
\newcommand{\mmP}[2][b]{\@imText[#1]{\Rot{#2}}}
\makeatother
\newcommand{\MMP}[2][b]{\mmP[#1]{#2}}

\makeatletter
\def\@pt{\ensuremath{\flat}}
\def\@om{\ensuremath{\sharp}}
\def\@comb#1#2{\ensuremath{{#1}^{#2}}}
\newcommand{\pTf}[1][f]{\@comb{#1}{\@pt}}
\newcommand{\pTX}[1][X]{\@comb{#1}{\@pt}}
\newcommand{\oMf}[1][f]{\@comb{#1}{\@om}}
\newcommand{\oMX}[1][X]{\@comb{#1}{\@om}}
\makeatother

\def\Zwei{2\!\!2}
\renewcommand{\upsilon}{\ensuremath{{\filterFont U}}}
\def\filterFont{\mathfrak}

\def\VauSenza{\ensuremath{\spaceFont{E}}}
\newcommand{\Vau}[1]{\ensuremath{\ensuremath{\VauSenza(#1)}}}

\newcommand{\dasI}[4][\mathcal{G}]{\ensuremath{\Omega_{#1}(#2 \mid #3,
    #4)}}

\def\AE{\Psi}
\def\Ge{{\cal G}}

\makeatletter
\newcommand{\@testOp}[2]{#1#2}
\newcommand{\posTest}[1][\phi]{\@testOp{#1}{?}}
\newcommand{\negTest}[1][\phi]{\@testOp{#1}{\text{\textquestiondown}}}
\makeatother
\usepackage{fancyhdr}
\def\EN{\end{document}}
\def\phi{\varphi}
\def\Zwei{2\!\!2}

\def\dasF{\textsf{F}\!\!{\textsf{F}}}
\renewcommand{\PowerSet}[1]{\ensuremath{\Zwei^{#1}}}
\renewcommand{\PowerSenza}{\ensuremath{\Zwei^{-}}}
\def\VauSenza{\ensuremath{\dasE}}
\renewcommand{\Vau}[1]{\ensuremath{\ensuremath{\VauSenza(#1)}}}
\renewcommand{\ProbSenza}{\$}
\renewcommand{\Prob}[1]{\ProbSenza\left(#1\right)}
\def\SubProb{\Prob}
\def\SubProbSenza{\ProbSenza}
\begin{document}
\maketitle\setcounter{tocdepth}{1}
\tableofcontents 
\begin{abstract}
  The stochastic interpretation of Parikh's game logic should not
  follow the usual pattern of Kripke models, which in turn are based
  on the Kleisli morphisms for the Giry monad, rather, a specific and
  more general approach to probabilistic nondeterminism is required. We
  outline this approach together with its probabilistic and measure
  theoretic basis, introducing in a leisurely pace the Giry monad and
  their Kleisli morphisms together with important techniques for
  manipulating them. Proof establishing specific techniques are given,
  and pointers to the extant literature are provided.

  After working through this tutorial, the reader should find it easier to
  follow the original literature in this and related areas, and it
  should be possible for her or him to appreciate measure theoretic arguments
  for original work in the areas of Markov transition systems, and
  stochastic effectivity functions.
\end{abstract}

\textbf{Computing Reviews Classification}: F.4.1, I.2.3, I.2.4, G.3
\onehalfspacing
\section{Objectives}
A minimal categorial framework is introduced in order to formulate
coalgebras and monads (which come here in their disguise as Kleisli
tripels). We specialize then to the category of measurable spaces,
discussing here in particular the Giry monad, with occasional side
glances to the upper closed functor.  This is complemented by a
discussion of morphisms for stochastic coalgebras (which will also be
used for the interpretation of various modal logics), giving among
others congruences, which will be put to use when discussing the
expressivity of Kripke models. But before doing so, we have a fairly
general look at bisimulations for various transition models, pointing
at difficulties arising for stochastic coalgebras. We expand then our
scenario by introducing stochastic effectivity functions, which we
will briefly investigate, and which will be used for an interpretation
of game logics. 

Not all topics can be treated in depth due to limitations in space,
but proofs are provided here and there, mostly for illustrating some
techniques. Two appendices are provided, one discusses a technical
device (the Souslin operation), the other one gives the important
$\pi$-$\lambda$-Theorem from Boolean $\sigma$-algebras.

The classic reference to coalgebras is the paper by
Rutten~\cite{Rutten}, the survey paper by
Venema~\cite{Venema-Handbook} focusses on representation issues, see
also~\cite{Jacobs-coalg}. The
present discussion is based
on~\cite{EED-Companion,EED-Alg-Prop_effFncts,EED-GameLogic-TR}. References
to extant pieces of literature are given when needed.

\section{Coalgebras}
Fix a category $\catC$ with an endofunctor $\funF$ (I assume that the
reader knows what a category is, and what a functor does).

\BeginDefinition{coalgebra}
An \emph{$\funF$-coalgebra} $(a, f)$\MMP{Coalgebra} over $\catC$ is an object $a$ of $\catC$
together with a morphism $f: a\to \funF a$. 
\EndDefinition

\BeginExample{power-set} Let $\catC$ be the category of sets with maps as morphisms, the functor is the power set functor $\PowerSenza$. $(A, f)$ is an $\PowerSenza$-coalgebra iff $f: A\to \PowerSet{A}$ is a map. This is in 1-1-correspondence with binary relations\MMP{Category of sets}:
\begin{equation*}
  \langle x, x'\rangle\in R \text{ iff } x'\in f(x).
\end{equation*}
Through this, \emph{transition systems} are studied. 
\EndExample

\BeginExample{automaton}
Let $X$ resp. $Y$ be the inputs and the outputs of an automaton\MMP{Automata} with
outputs. Define $\funF := (-\times Y)^{X}$ over the category of
sets. The $(A, f)$ is an $\funF$-coalgebra iff it is an automaton with
states $A$: Since $f: A\to (A\times Y)^{X}$, we have $f(a)(x)\in
A\times Y$, say, $f(a)(x) = \langle a', y\rangle$, hence $a'$ is the new state of the automaton, $y$ its output upon input $x$ in state $a$. 

Conversely, let $(X, Y, A, \delta)$ be an automaton with output, i.e., $\delta: A\times X\to A\times Y$ is a map. Currying gives a map $f: A\to (A\times Y)^{X}$ through $f(a)(x) := \delta(a, x)$. This means that $f$ is an $\funF$-coalgebra.
\EndExample

\BeginExample{trees}
Put $\funF A := \{*\}\cup A\times A$ with $*$ a new symbol\MMP{Trees}. $(A, f)$
is an $\funF$-coalgebra iff $f$ corresponds to a binary tree over
$A$. Put $f(a) := *$ iff $a$ is a leaf, and $f(a) = \langle
a_{1}, a_{2}\rangle$ iff $a_{1}$ and $a_{2}$ are offsprings of $a$. The tree may be infinite, though.
\EndExample

\medskip

Fix in what follows both $\catC$ and $\funF$.

\BeginDefinition{coalg-morph}
Let $(a_{1}, f_{1})$ and $(a_{2}, f_{2})$ be $\funF$-coalgebras. A
$\catC$-morphism $\phi: a_{1}\to a_{2}$ is a \emph{coalgebra morphism}
$(a_{1}, f_{1})\to (a_{2}, f_{2})$ iff $f_{2}\circ \phi =
\funF\phi\circ f_{1}$\MMP{Coalgebra morphism}. This means that the diagram
\begin{equation*}
\xymatrix{
a_{1}\ar[d]_{f_{1}}\ar[rr]^{\phi}&&a_{2}\ar[d]^{f_{2}}\\
\funF a_{1}\ar[rr]_{\funF \phi}&&\funF a_{2}
}
\end{equation*}
commutes.
\EndDefinition

\BeginProposition{coalg-is-cat}
$\funF$-coalgebras form a category with coalgebra morphisms as morphisms; composition is inherited from $\catC$. \QED
\EndProposition

\BeginExample{trans-syst}
Consider the coalgebras corresponding to transition systems from Example~\ref{power-set}. Then $\phi: (A_{1}, f_{1})\to (A_{2}, f_{2})$ is a coalgebra morphisms iff these conditions are satisfied:
\begin{enumerate}
\item $a_{1}'\in f_{1}(a_{1})$ implies $\phi(a_{1}')\in f_{2}(a_{1})$.
\item If $a_{2}'\in f_{2}(\phi(a_{1}))$, then there exists $a_{1}'\in f_{1}(a_{1})$ with $\phi(a_{1}') = a_{2}'$. 
\end{enumerate}

In fact, assume that $\phi$ is a coalgebra morphism. Then we note that $\PowerSet{\phi}(W) = \Bild{\phi}{W}$, and that we have this commuting diagram
\begin{equation*}
\xymatrix{
A_{1}\ar[d]_{f_{1}}\ar[rr]^{\phi}&&A_{2}\ar[d]^{f_{2}}\\
\PowerSet{A_{1}}\ar[rr]_{\PowerSet{\phi}}&&\PowerSet{A_{2}}
}
\end{equation*}
Assume that $a_{1}'\in f(a_{1})$, then 
\begin{equation*}
  \phi(a_{1}') \in \Bild{\phi}{f_{1}(a_{1})} = \PowerSet{\phi}(f_{1}(a_{1})) = f_{2}(\phi(a_{1})).
\end{equation*}
This gives us the first condition. On the other hand, let $a_{2}'\in f_{2}(\phi(a_{1})) = \PowerSet{\phi}(f_{1}(a_{1}))$, thus $a_{2}'\in \Bild{\phi}{f_{1}(a_{1})}$. But this implies that we can find $a_{1}'\in f_{1}(a_{1})$ with $\phi(a_{1}') = a_{2}'$. This provides us with the second condition. The converse direction offers itself as an exercise.

These morphisms are called \emph{bounded morphisms}\MMP{Bounded morphisms} in the theory of transition systems. In fact, reformulate $a'\in f_{1}(a)$ as $a\to_{1} a'$, similarly $b'\in f_2(b)$ as $b\to_{2}b'$. In this notation, the conditions above read
\begin{enumerate}
\item  $a_{1}\to_{1} a_{1}'$ implies $\phi(a_{1})\to_{2}\phi(a_{1}')$.
\item If $\phi(a_{1})\to_{2}a_{2}'$, then there exists $a_{1}'$ with $a_{1}\to_{1}a_{1}'$ such that $\phi(a_{1}') = a_{2}'$. 
\end{enumerate}
This indicates an interesting connection between coalgebras and transition systems.
\EndExample

\BeginExample{upper-closed}
Take the functor $\VauSenza$ of all upper closed subsets\MMP{Upper closed} of $\PowerSenza$. We did not yet define how $\VauSenza$ acts on maps, hence we have to transform $f: A\to B$ to $\Vau{f}: \Vau{A}\to \Vau{B}$. Let's see how to do this.

Let $\mathcal{G}\in\Vau{A}$, then $\mathcal{G}\subseteq\PowerSet{A}$ is upper closed. Thus the set 
$
\{H\subseteq B\mid \InvBild{f}{H}\in\mathcal{G}\}\subseteq \PowerSet{B}
$ is also upper closed (if $H_{1}\subseteq H_{2}$ and $\InvBild{f}{H_{1}}\in\mathcal{G}$, we note that $\InvBild{f}{H_{1}}\subseteq\InvBild{f}{H_{2}}$, and since $\mathcal{G}$ is upper closed, this implies that $\InvBild{f}{H_{2}}\in\mathcal{G}$). With this in mind, we put
\begin{equation*}
  \Vau{f}(\mathcal{G}) := \{H\subseteq B\mid \InvBild{f}{H}\in\mathcal{G}\}.
\end{equation*}
It can be shown that $\Vau{g\circ f} = \Vau{g}\circ \Vau{f}$ (see~\cite[2.3.14]{EED-Companion}). 

What do morphisms for $\VauSenza$ look like? Let's try:
\begin{equation*}
\xymatrix{
A_{1}\ar[d]_{f_{1}}\ar[rr]^{\phi}&&A_{2}\ar[d]^{f_{2}}\\
\Vau{A_{1}}\ar[rr]_{\Vau{\phi}}&&\Vau{A_{2}}
}
\end{equation*}
Let $\mathcal{H} = f_{2}(\phi(a_{1})) = \bigl(\Vau{\phi}\circ f_{1}\bigr)(a_{1})$, thus
\begin{equation*}
  H\in \mathcal{H} \text{ iff } H\in\Vau{\phi}\bigl(f_{1}(a_{1})\bigr)
\text{ iff }\InvBild{\phi}{H}\in f_{1}(a_{1}).
\end{equation*}
Consequently we have
\begin{equation*}
  f_{2}(\phi(a_{1})) = \{H\subseteq B \mid \InvBild{\phi}{H}\in f_{1}(a_{1})\}
\end{equation*}
as a qualifying condition for $\phi: A_{1}\to A_{2}$ to become a $\VauSenza$-morphism. 
\EndExample

\section{The $\SubProbSenza{}$-Functor}
Before entering the a discussion on the probability functor, we need
to know a little bit more about measurable spaces. Recall that a
measurable space $(X, \mathcal{A})$ is a set $X$ together with a
Boolean $\sigma$-algebra $\mathcal{A}$ on $X$, the $\sigma$ indicating
here that the Boolean algebra is closed under countable unions (and,
by implication, under countable intersections). 

\BeginDefinition{gen-sigma-alg}
Let $X$ be a set, $\mathcal{A}\subseteq\PowerSet{X}$ be a family of subsets of $X$. Then\MMP{$\sigma(\mathcal{A})$} 
\begin{equation*}
  \sigma(\mathcal{A}) := \bigcap\{\mathcal{B} \mid \mathcal{A}\subseteq \mathcal{B}, \mathcal{B} \text{ is a } \sigma-algebra\}
\end{equation*}
is the smallest $\sigma$-algebra on $X$ which contains $\mathcal{A}$. $\mathcal{A}$ is called a \emph{generator} of $\sigma(\mathcal{A})$. 
\EndDefinition

It is clear that $\sigma(\mathcal{A})$ is always a $\sigma$-algebra (check the properties). Also,  $\mathcal{A}\subseteq\PowerSet{X}$, the latter one being a $\sigma$-algebra. Note that $\sigma: \PowerSet{X}\to \PowerSet{X}$ is a closure operator, thus we have
\begin{itemize}
\item if $\mathcal{A}\subseteq\mathcal{B}$, then $\sigma(\mathcal{A})\subseteq\sigma(\mathcal{B})$,
\item $\mathcal{A}\subseteq\sigma(\mathcal{A})$,
\item $\sigma(\sigma(\mathcal{A})) = \sigma(\mathcal{A})$. In particular, $\sigma(\mathcal{A}) = \mathcal{A}$, whenever $\mathcal{A}$ is a $\sigma$-algebra. 
\end{itemize}

\BeginExample{Borel-sets}
The \emph{Borel sets $\Borel{\Real}$}\MMP{Borel sets} are defined as the smallest $\sigma$-algebra on $\Real$ which contains the open (or the closed) sets. We claim that 
\begin{equation*}
  \Borel{\Real} = \sigma\bigl(\bigl\{[r, +\infty[\mid r\in \Real\bigr\}\bigr) =: \mathcal{Q}.
\end{equation*}
In fact
\begin{itemize}
\item $\mathcal{Q}\subseteq\Borel{\Real}$, since 
  \begin{equation*}\textstyle
    [r, +\infty[ = \bigcap_{n\in \Nat}]r-\frac{1}{n}, \infty[,
  \end{equation*}
the latter sets are in $\Borel{\Real}$, since they are open.
\item $[r, s[\in \mathcal{Q}$ for $r<s$, since 
$
[r, s[ = [r, \infty[\setminus[s, \infty[.
$
\item $]r, s[ \in \mathcal{Q}$, since $]r, s[ = \bigcup_{n\in \Nat}[r+1/n, s[$.
\item All open sets in $\Real$ are in $\mathcal{Q}$, because each open set can be written as the union of countably many open intervals (and the open intervals are in $\mathcal{Q}$). 
\item All closed sets are in $\mathcal{Q}$ as well, since $\mathcal{Q}$ is closed under complementation. Hence $\Borel{\Real}\subseteq \mathcal{Q}$. 
\end{itemize}
\EndExample

The reals $\Real$ are \textbf{always} assumed to have the Borel $\sigma$-algebra. In what follows, we will usually write down measurable spaces without their $\sigma$-algebras, unless we have to. 

\BeginDefinition{meas-space}
Let $(X, \mathcal{A})$ and $(Y, \mathcal{B})$ be measurable spaces. A map $f: X\to  Y$ is called \emph{$\mathcal{A}$-$\mathcal{B}$-measurable}\MMP{Measurability} iff $\InvBild{f}{B}\in \mathcal{A}$ for all $B\in\mathcal{B}$. 
\EndDefinition

Note the similarity to continuity (\emph{inverse images of open sets
  are open} is the general definition), and to uniform continuity
(resp. \emph{inverse images of neighborhoods are neighborhoods}). Note
also that a measurable map is not necessarily a Boolean homomorphism
of the Boolean algebras $\mathcal{A}$ and $\mathcal{B}$ (constant maps
are always measurable, but are rarely homomorphisms of Boolean algebras). 

We should convince ourselves that we did indeed create a category.

\BeginProposition{meas-is-cat}
Measurable spaces with measurable maps form a category.
\EndProposition

\BeginProof
We need only to show: if $f: X\to Y$ and $g: Y\to Z$ are measurable,
so is $g\circ f: X\to Z$. Let $\mathcal{A}$, $\mathcal{B}$, and
$\mathcal{C}$ be the corresponding $\sigma$-algebras, then we obtain
from the measurability of $g$ that 
$\InvBild{g}{C}\in \mathcal{B}$ for
$C\in \mathcal{C}$, thus $\InvBild{f}{\InvBild{g}{C}} \in
\mathcal{A}$, since $f$ is also measurable. But $(g\circ f)^{-1} = f^{-1}\circ g^{-1}$, so the
assertion follows.  
\EndProof

\medskip

This is a helpful criterion for measurability, since it permits testing
only on a generator, which is usually more readily available than the
whole $\sigma$-algebra. 

\BeginLemma{meas-on-generators}
Let $(X, \mathcal{A})$ and $(Y, \mathcal{B})$ be measurable spaces, $f: X\to  Y$ a map, and assume that $\mathcal{B}$ is generated by $\mathcal{B}_{0}$. Then $f$ is $\mathcal{A}$-$\mathcal{B}$-measurable iff $\InvBild{f}{B}\in\mathcal{A}$ for all $B\in \mathcal{B}_{0}$. 
\EndLemma

\BeginProof
1.
The condition is clearly necessary for measurability, since $\mathcal{B}_{0}\subseteq \mathcal{B}$.

2.
The criterion is also sufficient. We show this through the
\emph{principle of good sets}\MMP{Principle of good sets} (see~\cite[Remark after Theorem
1.6.30]{EED-Companion}). It works like this. We want to show that
$\InvBild{f}{B}\in\mathcal{A}$ holds for all $B\in
\mathcal{B}$. Consider the set $\mathcal{G}$ of all ``good sets'',
\begin{equation*}
  \mathcal{G} := \{B\subseteq Y\mid \InvBild{f}{B}\in\mathcal{A}\}.
\end{equation*}
Then
\begin{enumerate}
\item $\mathcal{G}$ is a $\sigma$-algebra. This is so because $f^{-1}$
  is compatible with all the Boolean operations, e.g.,
  $\InvBild{f}{\bigcup_{i\in I}B_{i}} ) = \bigcup_{i\in I}\InvBild{f}{B_{i}}$.
\item $\mathcal{B}_{0}\subseteq\mathcal{G}$ by assumption. 
\end{enumerate}
Thus 
\begin{equation*}
  \sigma(\mathcal{B}_{0}) \subseteq \sigma(\mathcal{G}) = \mathcal{G}.
\end{equation*}
Hence $\mathcal{B} = \sigma(\mathcal{B}_{0}) \subseteq \mathcal{G}$,
but this means that $\InvBild{f}{B}\in\mathcal{A}$ for all
$B\in\mathcal{B}$. Hence $f$ is in fact $\mathcal{A}$-$\mathcal{B}$-measurable.
\EndProof

This is an easy consequence from Lemma~\ref{meas-on-generators}
together with Example~\ref{Borel-sets}:

\BeginLemma{real-meas}
A map $f: X\to \Real$ is measurable iff $\{x\in x\mid f(x) \geq r\}$
is a measurable subset of $X$ for all $r\in \Real$. \QED
\EndLemma

Note that we can replace the sets $\{f\geq r\}$ by $\{f \leq r\}$, by
$\{f> r\}$ or by $\{f<r\}$, since there are easy ways to compute one
through the other using only countable operations (such as
$\{f>r\} = \bigcup_{n\in \Nat}\{f\geq r+1/n\}$), and so on.

\BeginExample{char-fnct}
Let $\chi_{A}$ be the \emph{indicator function}\MMP{Indicator function} of $A\subseteq X$, thus
\begin{equation*}
  \chi_{A}(x) := \mathtt{if}\ x\in A\ \mathtt{then}\ 1\ \mathtt{else}\ 0\ \mathtt{fi}.
\end{equation*}
Then $\chi_{A}$ is a measurable function iff $A$ is a measurable
set. This becomes evident from 
\begin{equation*}
\{x\in X\mid \chi_{A}(x) \geq r\} =
  \begin{cases}
    \emptyset, & r > 1,\\
A, & 0< r \leq 1, \\
X, & r \leq 0
  \end{cases}
\end{equation*}
\EndExample

A measurable space induces a measurable structure on the space
\begin{equation*}
  \SubProb{X} := \{\mu \mid  \mu \text{ is a probability on (the
  $\sigma$-algebra of) } X\}
\end{equation*} in the following way. Define first
\begin{equation*}
  \beta(A, q) := \{\mu\in\SubProb{X} \mid \mu(A) \geq q\}
\end{equation*}
as the set of measures the value of which at event $A$ is not smaller than
$q$. 

\BeginDefinition{weak-star}
Given a measurable space $X$, its \emph{*-$\sigma$-algebra}\MMP{*-$\sigma$-algebra} is the smallest
$\sigma$-algebra on $\SubProb{X}$ which contains the sets $\{\beta(A,
q)\mid A\subseteq X\text{ measurable, } q\in \Real\}$.
\EndDefinition

Thus the *-$\sigma$-algebra is the smallest $\sigma$-algebra on
$\SubProb{X}$ which produces  measurable maps from the evaluation at events.

\BeginExample{dirac-kernel}
Define for the measurable space $X$ the embedding $\eta_{X}: X\to
\SubProb{X}$ upon setting $\eta_{X}(x)(A) := \chi_{A}(x)$. Then
$\eta_{X}$ is a measurable map. In fact, by
Lemma~\ref{meas-on-generators} and the definition of the
*-$\sigma$-algebra we have to show that the set 
\begin{equation*}
\InvBild{{\eta_{X}}}{\beta(A, q)} = \{x\in X\mid \eta_{X}(x)\in\beta(A, q)\}
\end{equation*}
is measurable for each measurable
set $A\subseteq X$ and each $q\in\Real$. But we have
\begin{equation*}
  \eta_{X}(x)\in\beta(A, q) \text{ iff }\chi_{A}(x)\geq q,
\end{equation*}
so that the set in question is measurable by
Example~\ref{char-fnct}. $\eta$ is usually called the \emph{Dirac
  kernel}, $\eta_{X}(x)$ the \emph{Dirac measure} on $x\in X$, which
is usually denoted by $\delta_{x}$, when the measurable space $X$ is
understood.

Reformulating, we see that $\eta_{X}: X\to \SubProbSenza{X}$ is a morphism in the category of
measurable spaces. 
\EndExample

We define now $\SubProbSenza$ for a measurable map and show that this
yields a measurable map again. This is the basis for
\begin{enumerate}
\item showing that $\SubProbSenza$ is an endofunctor on the category
  of measurable spaces,
\item establishing the properties of the Giry monad.
\end{enumerate}
Allora:

\BeginDefinition{act-dollar}
Let $X$ and $Y$ be measurable spaces, $f: X\to Y$ be a measurable
map. Define for $\mu\in\SubProb{X}$ and for $B\subseteq Y$ measurable
\begin{equation*}
  \SubProb{f}(\mu)(B) := \mu\bigl(\InvBild{f}{B}\bigr).
\end{equation*}
This is called the \emph{image measure}\MMP{Image measure} for $\mu$ under $f$.
\EndDefinition

The first step towards showing that $\SubProbSenza{}$ is an
endofunctor consists in showing that $\SubProbSenza$ transforms
measurable maps into measurable maps again (albeit on another space).

\BeginLemma{dollar-is-measurable}
Given $X$, $Y$ and $f$ as above, $\SubProb{f}: \SubProb{X}\to
\SubProb{Y}$ is measurable with respect to the *-$\sigma$-algebras. 
\EndLemma

\BeginProof
0.
We have to establish first that $(\SubProbSenza{f})(\mu)$ is a measure
on $Y$, provided $\mu\in\SubProbSenza{X}$. This is fairly
straightforward, let's have a look:
\begin{enumerate}
\item $\SubProb{f}(\mu)(\emptyset) = \mu(\InvBild{f}{\emptyset}) =
  \mu(\emptyset) = 0$, and $\SubProb{f}(\mu)(Y) = \mu(\InvBild{f}{Y})
  = \mu(X) = 1$.
\item Let $A$ and $B$ be disjoint measurable subsets of $Y$, then
  $\InvBild{f}{A}$ and $\InvBild{f}{B}$ are disjoint as well, thus
  \begin{align*}
    \SubProb{f}(\mu)(A\cup B) & = \mu(\InvBild{f}{A\cup B})\\
& = \mu(\InvBild{f}{A}\cup\InvBild{f}{B})\\
& = \mu(\InvBild{f}{A}) + \mu(\InvBild{f}{B})\\
& = \SubProb{f}(\mu)(A)+\SubProb{f}(\mu)(B).
  \end{align*}
\item If $\Folge{B}$ is an increasing sequence of measurable sets in
  $Y$ with $B := \bigcup_{n\in\Nat}B_{n}$, then
  $\bigl(\InvBild{f}{B_{n}}\bigr)_{n\in\Nat}$ is an increasing
  sequence of measurable subsets of $X$, and $\InvBild{f}{B}$ equals
  $\bigcup_{n\in\Nat}\InvBild{f}{B_{n}}$, thus
  \begin{equation*}
    \SubProb{f}(\mu)(B)  = \mu(\InvBild{f}{B})
 = \sup_{n\in\Nat}\mu(\InvBild{f}{B_{n}})
 = \sup_{n\in\Nat}\SubProb{f}(\mu)(B_{n}).
  \end{equation*}
\end{enumerate}

1.
We establish measurability by showing that the inverse image of a
generator to the *-$\sigma$-algebra in $Y$ is a *-$\sigma$-measurable
subset of $\SubProb{X}$. Then the assertion will follow from
Lemma~\ref{meas-on-generators}. In fact, let $B\subseteq Y$ be
measurable, then we claim that
\begin{equation*}
  \InvBild{(\SubProbSenza{f})}{\beta_{Y}(B, q)} =
  \beta_{X}(\InvBild{f}{B}, q)
\end{equation*}
holds. This is so because
\begin{align*}
  \mu\in\InvBild{(\SubProbSenza{f})}{\beta_{Y}(B, q)} 
& \Leftrightarrow
\SubProb{f}(\mu)\in\beta_{Y}(B, q)\\
& \Leftrightarrow
\SubProb{f}(\mu)(B) \geq q\\
& \Leftrightarrow
\mu(\InvBild{f}{B})\geq q\\
& \Leftrightarrow
\mu\in\beta_{X}(\InvBild{f}{B}, q).
\end{align*}
\EndProof

This yields as an immediate consequence

\BeginProposition{endofunctor}
$\SubProbSenza{}$ is an endofunctor on the category of measurable
spaces. \QED
\EndProposition

\section{The Giry Monad}

We want to determine the coalgebras for $\SubProbSenza{}$. Given a
measurable space $X$, a $\SubProbSenza$-coalgebra $(X, K)$ is a
measurable map $K: X\to \SubProbSenza{X}$. It will be necessary to
proceed a bit more general, and to characterize measurable maps $X\to
\SubProb{Y}$ first.

\BeginExample{dollar-coalgebras}
Let $X$ and $Y$ be measurable spaces, and $K: X\to \SubProb{Y}$ be a
measurable map (remember: $\SubProb{Y}$ carries the
*-$\sigma$-algebra). 
Thus
\begin{enumerate}
\item $K(x)$ is for every $x\in X$ a probability measure on (the
  measurable subsets of) $Y$.
\item Since $\InvBild{K}{\beta_{Y}(B, q)} = \{x\in X\mid K(x)(B)\geq q\}$,
  we see that the map $x\mapsto K(x)(B)$ is measurable for any fixed
  measurable set $B\subseteq Y$. 
\end{enumerate}
 Conversely, if we know that $x\mapsto K(x)(B)$ is measurable for any fixed
  measurable set $B\subseteq Y$, and that $K(x)$ is always a
  probability measure on $Y$, then the identity
  $$\InvBild{K}{\beta_{Y}(B, q)} = \{x\in X\mid K(x)(B)\geq q\}$$ shows that
  $K$ is a measurable map $X\to \SubProbSenza{Y}$. 

Thus we have in particular identified the \emph{coalgebras $(X, K)$ for the
$\SubProbSenza{}$-functor}\MMP{Coalgebras for 
$\SubProbSenza{}$} as maps $K: X\times\mathcal{A}\to [0, 1]$
such that
\begin{enumerate}
\item $K(x)$ is a probability measure on $X$,
\item $x\mapsto K(x)(A)$ is a measurable map for each measurable set $A$.
\end{enumerate}
(here $\mathcal{A}$ is the $\sigma$-algebra on $X$). $K$ is also known
in probabilistic circles as a \emph{Markov kernel} or a \emph{transition
probability}. In terms of transition systems: $K(x)(A)$ is the
probability of making a transition from $x$ to an element of the
measurable set $A$.
\EndExample

\BeginExample{morphisms-for-dollar}
We identify the morphisms for these coalgebras now. Let $(X, K)$ and
$(Y, L)$ be $\SubProbSenza{}$-coalgebras. A morphism $\phi$ for these
coalgebras must be a measurable map $\phi: X\to Y$, which is
compatible with the coalgebraic structure. This means in our case that
\begin{equation*}
\xymatrix{
X\ar[d]_{K}\ar[rr]^{\phi} && Y\ar[d]^{L}\\
\SubProbSenza{X}\ar[rr]_{\SubProbSenza{\phi}}&&\SubProbSenza{Y}
}
\end{equation*}
commutes. Thus
\begin{equation*}
 L(\phi(x))(B) = \bigl(L\circ \phi\bigr)(x)(B) =
 \bigl(\SubProb{\phi}\circ K\bigr)(x)(B)
= \SubProb{\phi}(K(x))(B)
= K(x)(\InvBild{\phi}{B}), 
\end{equation*}
meaning that the probability of hitting an element of $B$ from
$\phi(x)$ is the same as hitting an element of $\InvBild{\phi}{B}$
from $x$.
\EndExample

For proceeding further, we need the integral of a bounded measurable
function. Having this at our disposal, we can investigate the Giry
monad and put it into  context with other known monads by identifying common
properties. 

Fix for the moment a measurable space $X$ with $\sigma$-algebra
$\mathcal{A}$.

\BeginDefinition{meas-fncts}
Denote by $\MeasbFnct{X, \mathcal{A}} = \MeasbFnct{X}$ the set of all
bounded measurable functions on $X$. 
\EndDefinition

The algebraic structure of $\MeasbFnct{X}$ is easily identified.

\BeginLemma{is-vector-space}
$\MeasbFnct{X}$ is a real vector space with $\chi_{A}\in\MeasbFnct{X}$
iff $A\in\mathcal{A}$. 
\EndLemma

\BeginProof
0.  We know already from Example~\ref{char-fnct} that $\chi_{A}$
constitutes a measurable function iff the set $A$ is measurable.

1.
It is sufficient to show that $\MeasbFnct{X}$ is closed under addition
and under scalar multiplication. The latter property is fairly easy
established through Lemma~\ref{real-meas}, so let's try our hand on
the sum. We have for $r\in \Real$ and $f, g\in\MeasbFnct{X}$
\begin{align*}\textstyle
  \{x\in X\mid f(x)+g(x)<r\}
& = 
\bigcup_{q\in \Rational, q<r}\{x\in X\mid f(x)+g(x)<q\}\\
& =
\bigcup_{q\in \Rational, q<r}\Big(\bigcup_{a_{1}, a_{2}\in \Rational,
  a_{1}+a_{2}\leq q}\bigl(\{x\mid f(x)<a_{1}\}\cap\{x\mid g(x)<a_{2}\}\bigr)\Big)
\end{align*}
Because both $\{x\mid f(x)<a_{1}\}$ and $\{x\mid g(x)<a_{2}\}$ are
measurable sets, it follows that $ \{x\in X\mid f(x)+g(x)<r\}$ is a
measurable set, since $\Rational$ is
countable. The assertion now follows from Lemma~\ref{real-meas}.
\EndProof

It is also not difficult to show with the available tools that $\lim_{n\to \infty}f_{n}$ defines a
member of \MeasbFnct{X}, provided  $\Folge{f}\subseteq \MeasbFnct{X}$ such that $|f_{n}(x)|\leq B$ for all$n\in \Nat$ and $x\in X$, where $B\in \Real$ (the latter condition is necessary for making the limit a bounded function). 

\medskip

This permits us to define the integral of a bounded measurable
function. The elaborate process is somewhat technical and drawn out in
great detail in~\cite[Section 4.8]{EED-Companion}; we restrict
ourselves to presenting the result.

\BeginProposition{integral}
Let $\mu\in\SubProb{X}$ be a probability measure on $X$. There exists a unique map $\Phi: \MeasbFnct{X}\to \Real$
with these properties:
\begin{enumerate}
\item $\Phi(a\cdot f + b\cdot g) = a\cdot \Phi(f) + b\cdot
  \Phi(g)$, whenever $a, b\in \Real$ and $f, g\in \MeasbFnct{X}$ (linearity). 
\item $\Phi(f)\geq 0$, provided $f\geq0$, hence $\Phi$ is
  monotone (positivity).
\item $\Phi(\chi_{A}) = \mu(A)$ for all measurable sets
  $A\subseteq X$ (extension).
\item If $\Folge{f}$ is a sequence of bounded measurable
  functions such that the limit $\lim_{n\to \infty}f_{n}$ is bounded,
  then
  \begin{equation*}
    \Phi(\lim_{n\to \infty}f_{n})= \lim_{n\to \infty}\Phi(f_{n}).
  \end{equation*}
This is usually referred to as continuity.
\end{enumerate}
 \QED
\EndProposition

\textbf{Notation:} $\Phi(f)$ is written traditionally as\MMP{$\int_{X}f\ d\mu$} $\int_{X}f\ d\mu$, or as $\int_{X}
f(x)\ \mu(dx)$, if we want to emphasize the integration variable. We write $\int_{A}f\ d\mu$ for $\int_{X}f\cdot
\chi_{A}\ d\mu$. It is called the \emph{integral of $f$ with respect to
$\mu$}.

\medskip

\BeginExample{kleisli-map-1}
Let $K: X\to \SubProb{Y}$ be a measurable map for the measurable
spaces $X$ and $Y$. Define for $\mu\in\SubProb{X}$ 
\begin{equation*}
  K^{*}(\mu)(B) := \int_{X} K(x)(B)\ \mu(dx)
\end{equation*}
for $B\subseteq Y$ measurable\MMP{$K^{*}(\mu)$}. Then 
\begin{enumerate}
\item $K^{*}(\mu)\in\SubProbSenza{Y}$,
\item $\mu\mapsto K^{*}(\mu)$ is a measurable map $\SubProb{X}\to
  \SubProb{Y}$ with respect to the
  *-$\sigma$-algebras on $\SubProb{X}$ resp. $\SubProb{Y}$. 
\end{enumerate}
We establish only the first property. The second one is not
particularly difficult, but a bit more
time consuming to establish, so I refer you to~\cite[Example
2.4.8]{EED-Companion}. 

Let us have a look at the properties of a measure:
\begin{enumerate}
\item $K^{*}(\mu)(\emptyset) = \int_{X} 0\ d\mu = 0$ and $K^{*}(\mu)(Y)
  = \int_{X}1\ d\mu = \int_{X}\chi_{X}\ d\mu = \mu(X) = 1$.
\item Let $A$ and $B$ are disjoint measurable sets in $Y$, then 
  \begin{align*}
    K^{*}(\mu)(A\cup B) & = \int_{X}K(x)(A\cup B)\ \mu(dx)\\
& = \int_{X}\bigl(K(x)(A) + K(x)(B)\bigr)\ \mu(dx)&\text{ (each $K(x)$
  is a measure)}\\
& = \int_{X}K(x)(A)\ \mu(dx) + \int_{X}K(x)(B)\ \mu(dx)&\text{
  (additivity of the integral)}\\
& = K^{*}(\mu)(A) + K^{*}(\mu)(B).
  \end{align*}
\item Assume that $B_{1}\subseteq B_{2}\subseteq \dots \subseteq
  B_{n}\subseteq \dots$ is an increasing sequence of measurable sets
  in $Y$ with $B := \bigcup_{n\in\Nat}B_{n}$, then 
  \begin{equation*}
    K(x)(B) = \lim_{n\to \infty}K(x)(B_{n})
  \end{equation*}
for all $x\in X$, since each $K(x)$ is a measure, and $0\leq
K(x)(B)\leq 1$ for all $x\in X$, thus we obtain from continuity
\begin{align*}
  K^{*}(\mu)(B) & = \int_{X}K(x)(B)\ \mu(dx)\\
& = \int_{X}\lim_{n\to \infty}K(x)(B_{n})\ \mu(dx)\\
& = \lim_{n\to \infty}\int_{X}K(x)(B_{n})\ \mu(dx)\\
& = \lim_{n\to \infty}K^{*}(\mu)(B_{n}).
\end{align*}
\end{enumerate}
\EndExample

The next lemma shows an important technique for working with integrals, perceived as
extensions of measures; we will need this property badly when we are
discussing the Giry monad. The question arises naturally how to
integrate with respect to the measure $K^{*}(\mu)$, so we will try to
piece the integration together from the integrals wrt the measures
$K(x)$ for every $x\in X$, and  from the integral wrt $\mu$.

\BeginLemma{integral-repr}
Let $X$, $Y$, $K$, $\mu$ as above, then we have for all $f\in
\MeasbFnct{Y}$
\begin{enumerate}
\item $x\mapsto \int_{Y}f(y)\ K(x)(dy)$ defines a measurable and
  bounded function on $X$.
\item 
$ 
  \int_{X}f(x)\ K^{*}(\mu)(dx) = \int_{X}\bigl(\int_{Y}f(y)\
  K(x)(dy)\bigr)\ \mu(dx).
$ 
\end{enumerate}
\EndLemma

\BeginProof
0.
This is established very similar to the principle of good sets (see
the proof of Lemma~\ref{meas-on-generators}). Put
\begin{equation*}
  \mathcal{E} := \{f\in \MeasbFnct{Y}\mid \text{ the assertions are
    true for }f\}.
\end{equation*}
Then clearly $\mathcal{E}$ is a vector space over $\Real$. We show
that $\chi_{B}\in\mathcal{E}$ for $B\subseteq Y$ measurable; since
$\mathcal{E}$ is closed under bounded limits, the assertion follows
from the observation that linear combinations of indicator functions
are dense in $\MeasbFnct{X}$. The most complicated thing is to show
that $\chi_{B}\in\mathcal{E}$, which we will do now.

1.
We claim that both assertions are true for $f=\chi_{B}$, if $B\subseteq Y$ is a
measurable set. This is so because in this case we have for the first claim
\begin{equation*}
x\mapsto \int_{Y}f(y)\ K(x)(dy) = \int_{Y}\chi_{B}(y)\ K(x)(dy) = K(x)(B)
\end{equation*}
and $x\mapsto K(x)(B)$ defines a measurable function by definition of
the *-$\sigma$-algebra. From this we obtain
\begin{align*}
 \int_{X}\bigl(\int_{Y}\chi_{B}(y)\
  K(x)(dy)\bigr)\ \mu(dx) & = \int_{X}\bigl(K(x)(B)\bigr)\ \mu(dx)\\
& = K^{*}(\mu)(B)&\text{ (inner integral)}\\
& = \int_{X}\chi_{B}(y)\ K^{*}(\mu)(dy)&\text{ (extension property)}
\end{align*}

2.
If $f = \sum_{i=1}^{n}\alpha_{i}\cdot \chi_{B_{i}}$ is a 
step function with measurable sets $B_{1}, \dots, B_{n}$,
the assertion follows from the first part through the additivity of
the integral.

3.
The measurable step functions are dense in $\MeasbFnct{X}$ with
respect to pointwise convergence, so the assertion follows from the
second part, and from continuity of the integral. 
\EndProof

\medskip
This provides us with an amazing consequence. 

\BeginProposition{star-props}
Let $X$, $Y$ and $Z$ be measurable spaces, $K: X \to \SubProb{Y}$ and
$L: Y\to \SubProb{Y}$ be  measurable maps. Then 
$
  K^{*}\circ L^{*} = (K^{*}\circ L)^{*}.
$
\EndProposition

\BeginProof
0.
The proof is essentially a special case of Lemma~\ref{integral-repr}
(although it does not look like it), making use of the fact that
$y\mapsto K(y)(C)$ is a measurable map, and that $L(y)$ is a measure
for each $y\in Y$ and each $C\subseteq Z$ measurable.

1.
Let $\mu\in\SubProbSenza{X}$, and $C\subseteq Z$ be measurable, then
we have
\begin{align*}
  (L^{*}\circ K^{*})(\mu)(C) 
& = L^{*}\bigl(K^{*}(\mu)\bigr)(C)\\
& = 
\int_{Y}L(y)(C)\ K^{*}(\mu)(dy) &\text{ (definition of $-^{*}$)}\\
& = \int_{X}\bigl(\int_{Y}L(y)(C)\ K(x)(dy)\bigr)\ \mu(dx)&\text{
                                                            (apply
                                                            Lemma~\ref{integral-repr})}\\
&= \int_{X}L^{*}(K(x))(C)\ \mu(dx)& \text{ (note that
                                    }\bigl(\dots\bigr) =
                                    L^{*}(K(x))(C))\\
&= (L^{*}\circ K)^{*}(\mu)(C)&\text{ (definition of }-^{*})
\end{align*}
\EndProof

\BeginExample{lift-rel}
Lift $f: X\to \PowerSet{X}$ to $f^{*}: \PowerSet{X}\to \PowerSet{X}$
upon setting 
\begin{equation*}\textstyle
  f^{*}(A) := \bigcup_{x\in A}f(x).
\end{equation*}
Then an easy computation which the reader is invited to perform shows that $f^{*}\circ g^{*} = (f^{*}\circ
g)^{*}$ holds. 
\EndExample

\BeginExample{lift-upper}
Lift $f: X\to \Vau{X}$ to $f^{*}: \Vau{X}\to \Vau{X}$
upon setting 
\begin{equation*}
  f^{*}(\mathcal{C}) := \bigl\{B\subseteq X\mid \{x\mid B\in f(x)\}\in\mathcal{C}\bigr\}.
\end{equation*}
Then an easy computation with a similar scope shows that $f^{*}\circ g^{*} = (f^{*}\circ
g)^{*}$ holds holds. 
\EndExample

This is certainly not such a  strange coincidence.

\BeginDefinition{monad}
Let $\catC$ be a category, $\funT$ be a map which maps the objects in
$\catC$ to objects in $\catC$. Assume that we have a map $-^{*}$ which
maps morphisms $f: x\to \funT y$ to morphisms
$f^{*}:\funT x \to \funT y$ (called \emph{lifting}), and a morphism
$\eta_{x}: x\to \funT x$ (called \emph{embedding}) for each object $x$
in $\catC$. Then $(\funT, -^{*}, \eta)$ is called a \emph{monad}\MMP{Monad} iff
these conditions hold (here $x, y, z$ are objects in $\catC$):
\begin{enumerate}
\item\label{monad-1} $\eta_{x}^{*} = id_{\funT x}$.
\item\label{monad-2} $f^{*}\circ \eta_{x} = f$, whenever $f: x\to \funT y$.
\item\label{monad-3} $g^{*}\circ f^{*} = (g^{*}\circ f)^{*}$, whenever $f: x\to \funT
  y$, $g: y\to \funT z$. 
\end{enumerate}
\EndDefinition

What we call a monad here is usually called a \emph{Kleisli tripel} in
the literature. By Manes' Theorem~\cite[Theorem 2.4.4]{EED-Companion},
Kleisli tripels and monads are equivalent. Introducing monads in this
way has the advantage of not having to introduce natural
transformations and the slightly complicated diagrams associated with
it.

\medskip

Note that $\funT$ is only assumed to map objects to objects, but the
laws of a monad permit defining it on morphisms as well.

\BeginLemma{monad-to-endo}
Let $(\funT, -^{*}, \eta)$ be a monad over $\catC$. Then $\funT$ can
be extended to an endofunctor on $\catC$. 
\EndLemma

\BeginProof
Define $\funT f := (\eta_{y}\circ f)^{*}$ for $f: x\to y$. Then $\funT
f: \funT x \to \funT y$ is a morphism in $\catC$ with
\begin{enumerate}
\item $\funT id_{x} = \eta_{x}^{*} = id_{\funT x}$.
\item Assume $f: x\to y$ and $g: y\to z$, then we have
  \begin{align*}
    (\funT g)\circ (\funT f) & = (\eta_{z}\circ g)^{*}\circ (\eta_{y}\circ
    f)^{*}& \text{ (definition)}\\
& \stackrel{(\ddag)}{=} \bigl((\eta_{z}\circ g)^{*}\circ \eta_{y}\circ f\bigr)^{*}\\
& = (\eta_{z}\circ g\circ f)^{*}&\text{ (from~\ref{monad-3}. in Definition~\ref{monad})}\\
& = \funT (g\circ f).
  \end{align*}
Equation $(\ddag)$ uses the interplay of $\eta$ and the
$-^{*}$-operation, see property~\ref{monad-2}. in Definition~\ref{monad}.
\end{enumerate}
\EndProof

We are now in proud possession\MMP{Some monads}\label{proud} of the following monads:
\begin{enumerate}
\item Power set monad $\PowerSenza$, $f^{*}$ according to Example~\ref{lift-rel},
  $\eta_{X}(a) := \{a\}$ for $a\in X$.
\item Upper closed monad $\VauSenza$, $f^{*}$ according to
  Example~\ref{lift-upper}, $\eta_{X}(a) := \{A\subseteq X\mid a\in
  A\}$ for $a\in X$.
\item Probability monad $\SubProbSenza$, $K^{*}$ according to
  Example~\ref{kleisli-map-1}, $\eta$ is given through the Dirac
  kernel from Example~\ref{dirac-kernel}.  
\end{enumerate}

If you want to try your hand at other monads\MMP{Some other monads}, try these:
\begin{enumerate}
\item The \emph{ultra filter monad} over the category of sets. Let
  $U(X)$ be all ultrafilters over set $X$, and define
  $U(f): U(X)\to U(Y)$ for a given map $f: X\to Y$ verbatim as for
  upper closed sets in Example~\ref{upper-closed}, replacing the
  argument to $U(f)$ by an ultrafilter (it has to be shown that
  $U(f)(\mathcal{C})\in U(Y)$; this requires some thought). Define the embedding $X\to U(X)$ as in
  the case of the upper closed subsets.
\item The \emph{discrete probability monad} over the category of sets. Define
  \begin{equation*}\textstyle
    D(X) := \{p: X\to [0, 1]\mid p\text{ has countable support, and }
    \sum_{x\in X} p(x) = 1\},
  \end{equation*}
where the support of a map $p: X\to [0, 1]$ is defined as $\{x\in
X\mid p(x) \not= 0\}$ (hence the sum is defined). Let $f: X\to Y$ be a
map, $p\in D(X)$,  define 
\begin{equation*}
  D(f)(p)(y) := \sum_{f(x) = y}p(x).
\end{equation*}
Then show that $D(f)(p)\in D(Y)$. 
The embedding is defined as the Dirac kernel from
Example~\ref{dirac-kernel}. 
\end{enumerate}


We identify in what follows a monad with its functor, so that things
are a bit easier to handle. If, however, we need the components, we
will be explicit about them.

\BeginProposition{monad-yields-category}
A monad $\funT$ over category $\catC$ generates a new
category\MMP{Kleisli category}
$\catC_{\funT}$ in the following way:
\begin{enumerate}
\item The objects of $\catC_{\funT}$ are the objects of $\catC$,
\item A $\catC_{\funT}$-morphism $f: x\Trans y$\MMP{$f: x\Trans y$} in the new category is a $\catC$-morphism
  $f: x\to\funT y$ in $\catC$. 
\item The identity for $a$ in $\catC_{\funT}$ is $\eta_{a}: a \to
  \funT a$.
\item The composition $g \ast f$ of $f: x\Trans y$
  and $g: y\Trans z$ is defined through $g\ast f :=
  g^{*}\circ f$. 
\end{enumerate}
This category is called the \emph{Kleisli category} associated with $\funT$
(and $\catC$, of course).
\EndProposition

\BeginProof
We have to show that the laws of a category are satisfied, hence in
particular that
Kleisli composition is associative. In fact, we
have 
\begin{align*}
  (h\ast g)\ast f & = (h\ast g)^{*}\circ f \\
& = (h^{*}\circ g)^{*}\circ f &&\text{(definition of $h\ast g$)}\\
& = h^{*}\circ g^{*}\circ f && \text{(property 3. in a monad)}\\
& = h^{*}\circ (g\ast f)&&\text{(definition of $g\ast f$)}\\
& = h\ast (f\ast g)
\end{align*}
The laws for the identity are easily checked from the properties of
$-^{*}$. Thus $\catC_{\funT}$ is indeed a category. 
\EndProof

\BeginExample{kleisli-for-power}
The Kleisli morphisms for the power set monad are exactly the
relations, and we have for $R: X\Trans Y$ and $S: Y\Trans Z$ that
\begin{equation*}
  z \in (R\ast S)(x) \Leftrightarrow z\in S(y) \text{ for some }y\in R(x).
\end{equation*}
This is immediate. 
\EndExample

\BeginExample{kleisli-for-prob}
The Kleisli morphisms for the Giry monad are exactly the stochastic
relations, a.k.a. transition
probabilities. Let $K: X\Trans Y$ and $L: Y\Trans Z$ be stochastic
relations, then we have 
\begin{equation*}
  (L\ast K)(x)(C) = \int_{Y}L(y)(C)\ K(x)(dy),
\end{equation*}
when $x\in X$ and $C\subseteq Z$ is a measurable set. This follows
immediately from the definition of $-^{*}$ for this monad.
\EndExample

\section{Playing Around with Morphisms}
\label{sec:play-with-morph}

Measurable spaces form a category under measurable maps. In fact, given
a measurable space $(X,\mathcal{A})$, 
\begin{enumerate}
\item we can find for a map $f: Z\to X$ a smallest $\sigma$-algebra
  $\mathcal{C}$ on $Z$ which renders $f$ a $(Z, \mathcal{C})$-$(X,
  \mathcal{A})$-measurable map. Take simply 
  \begin{equation*}
    \mathcal{C} := \{\InvBild{f}{A}\mid A\in\mathcal{A}\}.
  \end{equation*}
$\mathcal{C}$ is called the \emph{initial $\sigma$-algebra}\MMP{Initial} with respect to
$f$ and $\mathcal{A}$. 
\item we can find for a map $g: X\to Y$ a largest $\sigma$-algebra
  $\mathcal{B}$ on $Y$ such that $g$ is $\mathcal{A}$-$(Y,
  \mathcal{B})$-measurable. Take simply
  \begin{equation*}
    \mathcal{B} := \{B\subseteq Y\mid \InvBild{f}{B}\in \mathcal{A}\}.
  \end{equation*}
$\mathcal{B}$ is called the \emph{final $\sigma$-algebra}\MMP{Final} with respect to
$g$ and $\mathcal{A}$. 
\end{enumerate}
Both constructions extend easily to families of maps $f_{i}: Z_{i}\to
X$ resp. $g_{i}: X\to Y_{i}$. For example, the *-$\sigma$-algebra on
$\SubProb{X}$ is
the initial $\sigma$-algebra with respect to the family $\{ev_{A}\mid
A\in \mathcal{A}\}$, with $ev_{A}: \mu\mapsto \mu(A)$, and the
\emph{product-$\sigma$-algebra} $\mathcal{A}\otimes\mathcal{B}$\MMP{$\mathcal{A}\otimes\mathcal{B}$} on the
Cartesian product $X\times Y$ is the initial $\sigma$-algebra on
$X\times Y$ with respect to the projections $\pi_{X}$ and $\pi_{Y}$. 

Given an equivalence relation $\tau$ on $X$, define
$\Faktor{\mathcal{A}}{\tau}$ as the final $\sigma$-algebra on the set
$\Faktor{X}{\tau}$ of equivalence classes with respect to the factor
map $\rho_{\tau}: x \mapsto \Klasse{x}{\tau}$. We assume that this
space is always equipped with this $\sigma$-algebra. 

Equivalence relation $\tau$ defines also a $\sigma$-algebra on $X$,
the $\sigma$-algebra of \emph{$\tau$-invariant (measurable) sets},
upon setting\MMP{$\Sigma_{\tau}$}
\begin{equation*}
  \Sigma_{\tau} := \Sigma_{\tau, \mathcal{A}} := \{A\in
  \mathcal{A}\mid A\text{ is $\tau$-invariant}\}
\end{equation*}
(recall that set $A$ is \emph{$\tau$-invariant} iff it is the union of
$\tau$-classes, or, equivalently, iff $x\in A$ and
$\isEquiv{x}{x'}{\tau}$ together imply $x'\in A$). Look at this equivalence
\begin{equation*}
  A\in \Sigma_{\tau}\Leftrightarrow \Bild{\rho_{\tau}}{A}\in \Faktor{\mathcal{A}}{\tau}.
\end{equation*}
Does it always hold? In fact, this is true, and it hinges on the
equality $\InvBild{\rho_{\tau}}{\Bild{\rho_{\tau}}{A}} = A$ for
$\tau$-invariant $A\subseteq X$ ($\subseteq:$ If $\rho_{\tau}(x) \in
\Bild{\rho_{\tau}}{A}$, there exists $x'\in A$ with
$\isEquiv{x}{x'}{\rho}$, hence $x\in A$; $\supseteq:$ is
trivial). An equivalent formulation is evidently
\begin{equation*}
  \Faktor{\mathcal{A}}{\tau} = \{A\subseteq \Faktor{X}{\tau}\mid \InvBild{\rho_{\tau}}{A}\in\Sigma_{\tau}\}.
\end{equation*}

$\Sigma_{\tau}$ is
a fairly important $\sigma$-algebra, as we will see. Occasionally one
considers as an equivalence relation the \emph{kernel}
$\Kern{f}$\MMP{Kernel, $\Kern{f}$} of a measurable map $f: X\to Y$. It is
defined as
\begin{equation*}
  \Kern{f} := \{\langle x, x'\rangle\mid f(x) = f(x')\}.
\end{equation*}

We are now in a position to define congruences for stochastic
relations on a measurable space $X$. Remember that a stochastic
relation $K: X\Trans X$ is a coalgebra $(X, K)$ for the Giry functor.

\BeginDefinition{factor-relation}
Let $K: X\Trans X$ be a stochastic relation. An equivalence
relation $\tau$ is called a \emph{congruence for $K$} iff there
exists a stochastic relation $K_{\tau}: \Faktor{X}{\tau}\Trans
\Faktor{X}{\tau}$ such that $\rho_{\tau}: K\to K_{\tau}$ is a
morphism. 
\EndDefinition

Let us see what it means that $\tau$ is a congruence for $K$. Since we
are dealing with coalgebras here, this means that this diagram
commutes
\begin{equation*}
\xymatrix{
X\ar[d]_{K}\ar[rr]^{\rho_{\tau}}&&\Faktor{X}{\tau}\ar[d]^{K_{\tau}}\\
\SubProb{X}\ar[rr]_{\SubProb{\rho_{\tau}}}&& \SubProb{\Faktor{X}{\tau}}
}
\end{equation*}
Hence we have for $x\in X$ and $A\subseteq \Faktor{X}{\tau}$
measurable this equality
\begin{equation*}
  K_{\tau}(\Klasse{x}{\tau})(A) = (K_{\tau}\circ
  \rho_{\tau})(x)(A)
= \bigl(\SubProb{\rho_{\tau}}\circ K(x)\bigr)(A) = K(x)(\InvBild{\rho_{\tau}}{A}).
\end{equation*}
This means that the behavior of $K(x)$ on the $\sigma$-algebra
$\Sigma_{\tau}$ determines the behavior of
$K_{\tau}(\Klasse{x}{\tau})$ completely. This is intuitively somewhat
satisfying: if $\tau$ cannot distinguish between $x$ and $x'$, then
$K(x)(A)$ should be the same as $K(x')(A)$ for all $A$ the elements of
which $\tau$ cannot tell apart (actually, this is how a congruence was
first defined for $K$). 

In universal algebra there is a strong connection between the kernels
of morphisms and congruences, actually, e.g., in Abelian groups, the kernel of a morphism is a
congruence, and vice versa. In general, additional conditions are
necessary. A measurable map $f: X\to Y$ is called \emph{strong}\MMP{Strong} iff $f$ is
surjective so that $Y$ carries the final $\sigma$-algebra with respect
to $f$; note that being strong is an intrinsic property of $f$ and is
independent of any $\SubProbSenza$-coalgebra.

\BeginProposition{congruence-vs-kernelhom}
Let $(X, K)$ and $(Y, L)$ be $\SubProbSenza$-coalgebras, and $f: (X,
K)\to (Y, L)$ is a strong morphism. Then $\Kern{f}$ is a
congruence. Conversely, if $\tau$ is a congruence for $(X, K)$, then
$\rho_{\tau}$ is a strong morphism. 
\EndProposition

\BeginProof
The assertion about $\rho_{\tau}$ is trivial from the construction,
and since $\Kern{\rho_{\tau}} = \tau$. The converse follows from some
observations on general coalgebras based on sets, and a characterization of
$\Sigma_{f}$ for strong $f$ in~\cite[Section 2.6.2]{EED-Companion}.
\EndProof 

We want to define subsystems for a $\SubProbSenza$-coalgebra $(X, K)$,
$X = (X, \mathcal{A})$ being a measurable space again. Subsystems are determined through a
sub-$\sigma$-algebra $\mathcal{B}\subseteq\mathcal{A}$ and through a
transition law, say, $L$. Note that the identity $i_{X}: X\to X$ is
$\mathcal{A}$-$\mathcal{B}$-measurable iff
$\mathcal{B}\subseteq\mathcal{A}$. This time we have to make the
$\sigma$-algebra explicit.

\BeginDefinition{subsystem}
$\bigl((X, \mathcal{B}), L\bigr)$ is a \emph{subsystem}\MMP{Subsystem} of $\bigl((X,
\mathcal{A}), K\bigr)$ iff the identity is a morphism $i_{X}: \bigl((X, \mathcal{A}), K\bigr)\to \bigl((X, \mathcal{B}), L\bigr)$.
\EndDefinition

Again, we have this diagram, which commutes for a subsystem:
\begin{equation*}
\xymatrix{
(X, \mathcal{A})\ar[d]_{K}\ar[rr]^{i_{X}}&&(X, \mathcal{B})\ar[d]^{L}\\
\SubProb{X, \mathcal{A}}\ar[rr]_{\SubProb{i_{X}}}&& \SubProb{X, \mathcal{B}}
}
\end{equation*}
Hence $K(x)(B) = L(x)(B)$ for all $x\in X$, and all
$B\in \mathcal{B}$, which may be interpreted either that $K(x)$ is an
extension to $L(x)$ or that $L(x)$ is the restriction of $K(x)$,
depending on the situation at hand. Sometimes a subsystem is called a
\emph{state bisimulation}, but I think that this is an unfortunate
name, because bisimilarity as a means of comparing the expressivity
of systems through a mediator is nowhere to be seen. A subsystem will
be identified it through its $\sigma$-algebra $\mathcal{B}$; the
coalgebra is then defined through the restriction to $\mathcal{B}$.

It is immediate that a congruence $\tau$ defines a subsystem with
$\Sigma_{\tau}$ as the defining $\sigma$-algebra. 

\section{Bisimulations}
\label{sec:bisimulations}

The notion of bisimilarity is fundamental for the application of
coalgebras to system modelling. Bisimilar coalgebras behave in a
similar fashion, witnessed by a mediating system.

\BeginDefinition{bisim-coalg}
Let $\funF$ be an endofunctor on a category $\catC$. The
$\funF$-coalgebras $(a, f)$ and $(b, g)$ are said to be \emph{bisimilar} iff
there exists a coalgebra $(m, v)$ and coalgebra
morphisms\MMP{Bisimilar, mediating} 
$$
\xymatrix{
  (a, f)&(m, v)\ar[l]\ar[r]&(b, g).
}
$$
The coalgebra $(m, v)$ is called \emph{mediating}. 
\EndDefinition
Thus we obtain this characteristic diagram with $\ell$ and $r$ as the
corresponding morphisms.
\begin{equation*}
\xymatrix{
a\ar[d]_{f} && m\ar[d]^{v}\ar[ll]_{\ell}\ar[rr]^{r} && b\ar[d]^{g}\\
\funF a && \funF m\ar[ll]^{\funF \ell}\ar[rr]_{\funF r} && \funF b
}
\end{equation*}
This gives us 
$f\circ \ell  = (\funF\ell)\circ v \text{ and }$ together with $g\circ r  = (\funF r)\circ v.$
It is easy to see why $(M, m)$ is called mediating. 

\medskip

Bisimilarity was originally investigated when concurrent systems
became of interest~\cite{Hennessy-Milner}. The original formulation, however, was not
coalgebraic but rather relational. Here it is (for the sake of easier
reading, we use arrows rather that relations or maps into the
respective power set):

\BeginDefinition{bisim-relat}
Let $(S, \Trans_{S})$ and $(T, \Trans_{T})$ be transition
systems. Then $B\subseteq S\times T$ is called a
\emph{bisimulation}\MMP{Bisimulation, $\PowerSenza$} iff for
all $\langle s, t\rangle\in B$ these conditions are satisfied:
\begin{enumerate}
\item if $s\Trans_{S}s'$, then there is a $t'\in T$ such that
  $t\Trans_{T}t'$ and $\langle s', t'\rangle\in B$,
\item if $t\Trans_{T}t'$, then there is a $s'\in S$ such that
  $s\Trans_{S}s'$ and $\langle s', t'\rangle\in B$.
\end{enumerate}
\EndDefinition
Hence a bisimulation simulates transitions in one system through the
other one. On first sight, these notions of bisimilarity are not
related to each other. Recall that transition systems are coalgebras
for the power set functor $\PowerSenza$. This is the connection:

\BeginTheorem{ascel-bisim}
Given the  transition systems $(S, \Trans_{S})$ and $(T, \Trans_{T})$
with the associated $\PowerSenza$-coalge\-bras $(S, f)$ and $(T, g)$, then
these statements are equivalent for $B\subseteq S\times T$\MMP{Aczel's
Theorem}: 
\begin{enumerate}
\item\label{ascel-bisim-1} $B$ is a bisimulation.
\item\label{ascel-bisim-2} There exists a $\PowerSenza$-coalgebra structure $h$ on
  $B$ such that 
$
\xymatrix{
  (S, f)&(B, h)\ar[l]\ar[r]&(T, g)
}
$
with the projections as morphisms is mediating. 
\end{enumerate}
\EndTheorem

\BeginProof
That 
$
\xymatrix{
  (S, f)&(B, h)\ar[l]_{\pi_{S}}\ar[r]^{\pi_{T}}&(T, g)
}
$
is mediating follows from commutativity of this diagram.
\begin{equation*}
\xymatrix{
S\ar[d]_{f} && B\ar[d]_{h}\ar[ll]_{\pi_{S}}\ar[rr]^{\pi_{T}} && T\ar[d]^{g}\\
\PowerSenza(S) && \PowerSenza(B)\ar[ll]^{\PowerSenza(\pi_{S})}\ar[rr]_{\PowerSenza(\pi_{T})}&&\PowerSenza(T)
}
\end{equation*}

``\labelImpl{ascel-bisim-1}{ascel-bisim-2}'': We have to construct a map
$h: B\to \PowerSenza(B)$ such that
$
  f(\pi_{S}(s, t))  = \PowerSenza(\pi_{S})(h(s, t))
$
and
$
f(\pi_{T}(s, t)) = \PowerSenza(\pi_{T})(h(s, t))
$
for all $\langle s, t\rangle \in B$. The choice is somewhat obvious:
put for $\langle s, t\rangle \in B$
\begin{equation*}
  h(s, t) := \{\langle s', t'\rangle \in B \mid s\Trans_{S}s', t\Trans_{T}t'\}.
\end{equation*}
Thus $h: B\to \PowerSenza(B)$ is a map, hence $(B, h)$ is a
$\PowerSenza$-coalgebra. 

Now fix $\langle s, t\rangle\in B$, then we claim that $f(s) =
\PowerSenza(\pi_{S})(h(s, t)).$

\begin{description}
\item[``$\subseteq$'':] Let $s'\in f(s)$, hence
  $s\Trans_{S}s'$, thus there exists $t'$ with
  $\langle s', t'\rangle\in B$ such that $t\Trans_{T}t'$, hence
  \begin{equation*}
    s'   \in \{\pi_{S}(s_{0}, t_{0}) \mid \langle s_{0}, t_{0}\rangle \in
         h(s, t)\} 
         = \{s_{0}\mid \langle s_{0}, t_{0}\rangle \in h(s, t)\text{ for some
         $t_{0}$}\} 
         = \PowerSenza(\pi_{S})(h(s, t)).
  \end{equation*} 
\item[``$\supseteq$'':] If $s'\in\PowerSenza(\pi_{S})(h(s, t))$, then in
  particular $s\Trans_{S}s'$, thus $s'\in f(s)$.
\end{description}

Thus we have shown that $\PowerSenza(\pi_{S})(h(s, t)) = f(s) =
f(\pi_{S}(s, t))$. One shows $\PowerSenza(\pi_{T})(h(s, t)) = g(t) =
f(\pi_{T}(s, t))$ in exactly the same way. We have constructed
$h$ such that $(B, h)$ is a $\PowerSenza$-coalgebra, and such that the
diagrams above commute.

``\labelImpl{ascel-bisim-2}{ascel-bisim-1}'': Assume that $h$ exists
with the properties described in the assertion, then we have to show
that $B$ is a bisimulation. Now let $\langle s, t\rangle \in B$ and
$s\Trans_{S}s'$, hence $s'\in f(s) = f(\pi_{S}(s, t)) =
\PowerSenza(\pi_{S})(h(s, t))$. Thus there exists $t'$ with $\langle
s', t'\rangle \in h(s, t)\subseteq B$, and hence $\langle s',
t'\rangle \in B$. We claim that $t\Trans_{T}t'$, which is tantamount
to saying $t'\in g(t)$. But $g(t) = \PowerSenza(\pi_{T})(h(s, t))$, and
$\langle s', t'\rangle \in h(s, t)$, hence $t'\in
\PowerSenza(\pi_{T})(h(s, t)) = g(t)$. This establishes
$t\Trans_{T}t'$. A similar argument finds $s'$ with $s\Trans_{S}s'$
with $\langle s', t'\rangle\in B$ in case $t\Trans_{T}t'$.

This completes the proof.
\EndProof

Thus we may use bisimulations for transition systems as relations and
bisimulations as coalgebras interchangeably, and this characterization
suggests a definition in purely coalgebraic terms for those cases in
which a set-theoretic relation is not available or not adequate. The connection to
$\PowerSenza$-coalgebra morphisms and bisimulations is further
strengthened by investigating the graph of a morphism (recall that the
graph\MMP{$\Graph{r}$} of a map $r: S\to T$ is the relation $\Graph{r} := \{\langle s,
r(s)\rangle \mid s\in S\}$). 

\BeginProposition{morph-as-bisim}
Given coalgebras $(S, f)$ and $(T, g)$ for the power set functor
$\PowerSenza$, $r: (S, f)\to (T, g)$ is a morphism iff $\Graph{r}$ is a
bisimulation for $(S, f)$ and $(T, g)$. 
\EndProposition

\BeginProof
1.
Assume that $r: (S, f)\to (T, g)$ is a morphism, so that $g\circ r =
\PowerSenza(r)\circ f$. Now define 
\begin{equation*}
  h(s, t):=\{\langle s', r(s')\rangle \mid s'\in f(s)\}\subseteq \Graph{r}
\end{equation*}
for $\langle s, t\rangle \in \Graph{r}$. Then
$g(\pi_{T}(s, t)) = g(t) = \PowerSenza(\pi_{T})(h(s, t))$ for $t = r(s)$.

\begin{description}
\item[``$\subseteq$'':] If $t'\in g(t)$ for $t = r(s)$, then
  \begin{equation*}
    \begin{array}{lll}
    t'\in g(r(s))  & = \PowerSenza(r)(f(s))
                   & = \{r(s') \mid s'\in f(s)\}\\
                   &= \PowerSenza(\pi_{T})(\{\langle s', r(s')\rangle \mid s'\in
                    f(s)\})
                   &= \PowerSenza(\pi_{T})(h(s, t))
    \end{array}
  \end{equation*}
\item[``$\supseteq$'':] If $\langle s', t'\rangle \in h(s, t)$, then
  $s'\in f(s)$ and $t' = r(s')$, but this implies
  $t'\in \PowerSenza(r)(f(s)) = g(r(s)).$
\end{description}

Thus $g\circ \pi_{T}= \PowerSenza(\pi_{T})\circ h$. The equation 
$f\circ \pi_{S} = \PowerSenza(\pi_{S})\circ h$ is established similarly.

Hence we have found a coalgebra structure $h$ on $\Graph{r}$ such that
\begin{equation*}
\xymatrix{
(S, f)&&(\Graph{r}, h)\ar[ll]_{\pi_{S}}\ar[rr]^{\pi_{T}}&&(T, g)
}
\end{equation*}
are coalgebra morphisms, so that $(\Graph{r}, h)$ is now officially a bisimulation. 

2.
If, conversely, $(\Graph{r}, h)$ is a bisimulation with the projections
as morphisms, then we have $r = \pi_{T}\circ \pi_{S}^{-1}$. Then
$\pi_{T}$ is a morphism, and $\pi_{S}^{-1}$ is a morphism as well
(note that we work on the graph of $r$). So $r$ is a morphism. 
\EndProof

{
\def\funV{\VauSenza}
Let us have a look at  upper closed sets. There we find a comparable
situation. We cannot, however, translate the definition directly, because we do
not have access to the transitions proper, but rather to the sets from
which the next state may come from. Let  $(S, f)$ and
$(T, g)$ be $\funV$-coalgebras, and assume that $\langle s,
t\rangle\in B$. Assume $X\in f(s)$, then we want to find $Y\in g(t)$
such that, when we take $t'\in Y$, we find a state $s'\in X$
with $s'$ being related via $B$ to $s'$, and vice versa. Formally: 
\BeginDefinition{bisim-upperclosed}
Let again
\begin{equation*}
  \funV S := \{V\subseteq\PowerSenza(S) \mid V\text{ is upper closed}\}
\end{equation*}
be the endofunctor on $\catSET$ which assigns to set $S$ all upper
closed subsets of $\PowerSenza S$\MMP{Bisimulation, $\funV$}. Given $\funV$-coalgebras $(S, f)$
and $(T, g)$, a subset $B\subseteq S\times T$ is called a
\emph{\index{bisimulation}bisimulation} of $(S, f)$ and $(T, g)$ iff for each $\langle s, t\rangle \in B$
\begin{enumerate}
  \item for all $X\in f(s)$ there exists $Y\in g(t)$ such that for
    each $t'\in Y$ there exists $s'\in X$ with $\langle s', t'\rangle \in B$,
  \item for all $Y\in g(t)$ there exists $X\in f(s)$ such that for
    each $s'\in X$ there exists $t'\in Y$ with $\langle s', t'\rangle \in B$.
\end{enumerate}
\EndDefinition

We have then a comparable characterization of bisimilar coalgebras~\cite{EED-Game-Coalg}.

\BeginProposition{bisim-through-morph}
Let $(S, f)$ and $(T, g)$ be coalgebras for $\funV$. Then the
following statements are equivalent for $B\subseteq S\times T$ with
$\Bild{\pi_{S}}{B} = S$ and $\Bild{\pi_{T}}{B} = T$
\begin{enumerate}
  \item \label{bisim-is-coalg-1} $B$ is a
    bisimulation of $(S, f)$ and $(T, g)$.
  \item \label{bisim-is-coalg-2} There exists a coalgebra structure
    $h$ on $B$ so that the projections $\pi_S: B\to S, \pi_T: B \to T$
  are morphisms  
  $
  \xymatrix{
  (S, f)&(B, h)\ar[l]_{\pi_S}\ar[r]^{\pi_T}&(T, g).
  }
  $
\end{enumerate}
\EndProposition

\BeginProof
``\labelImpl{bisim-is-coalg-1}{bisim-is-coalg-2}'': 
Define $\langle s, t\rangle\in B$
\begin{equation*}
h(s, t) := \{D \subseteq B \mid \Bild{\pi_S}{D}\in f(s)\text{ and }\Bild{\pi_T}{D}\in f(t)\}.
\end{equation*}
Hence $h(s, t)\subseteq\PowerSet{S}$, and because both $f(s)$ and  $g(t)$ are upper
closed, so is $h(s, t)$. 

Now fix $\langle s, t\rangle\in B$. We show first that 
$
f(s) = \{\Bild{\pi_{S}}{Z}\mid Z\in h(s, t)\}.
$
From the definition of $h(s, t)$ it follows that $\Bild{\pi_{S}}{Z}\in
f(s)$ for each $Z\in h(s, t)$. So we have to establish the other
inclusion. Let $X\in f(s)$, then 
$
X = \Bild{\pi_S}{\InvBild{\pi_S}{X}},
$
because $\pi_{S}: B\to S$ is onto, so it suffices to show that $\InvBild{\pi_S}{X}\in h(s, t)$, hence that
$
\Bild{\pi_T}{\InvBild{\pi_S}{X}}\in g(t).
$
Given $X$ there exists $Y\in g(t)$ so that
for each $t'\in Y$ there exists $s'\in X$ such that $\langle s', t'\rangle \in B$. 
Thus 
$
Y = \Bild{\pi_T}{(X\times Y)\cap B}.
$
But this implies
$
Y \subseteq \Bild{\pi_T}{\InvBild{\pi_S}{X}},
$
hence 
$
Y\subseteq\Bild{\pi_T}{\InvBild{\pi_S}{X}} \in g(t).
$
One similarly shows that $g(t) = \{\Bild{\pi_{T}}{Z}\mid Z\in h(s,
t)\}$. 

In a second step, we show that 
\begin{equation*}
\{\Bild{\pi_{S}}{Z}\mid Z \in h(s, t)\} = \{C \mid
\InvBild{\pi_{S}}{C}\in h(s, t)\}.
\end{equation*}
In fact, if $C = \Bild{\pi_{S}}{Z}$ for some $Z\in h(s, t)$, then $Z
\subseteq \InvBild{\pi_{S}}{C} =
\InvBild{\pi_{S}}{\Bild{\pi_{S}}{Z}}$, hence $\InvBild{\pi_{S}}{C}\in
h(s, t)$. If, conversely, $Z := \InvBild{\pi_{S}}{C}\in h(s, t)$, then
$C = \Bild{\pi_{S}}{Z}$. Thus we obtain 
\begin{equation*}
  f(s)  = \{\Bild{\pi_{S}}{Z}\mid Z\in h(s, t)\}
 = \{C \mid \InvBild{\pi_{S}}{C}\in h(s, t)\}
 = (\funV \pi_{S})(h(s, t))
\end{equation*}
for $\langle s, t\rangle \in
B$. 
Summarizing,  this means that 
$
\pi_S: (B, h) \to (S, f)
$
is a morphism. A very similar argumentation shows that  
$
\pi_T : (B, h) \to (T, g) 
$
is a morphism as well. 

``\labelImpl{bisim-is-coalg-2}{bisim-is-coalg-1}'': Assume, conversely, that
the projections are coalgebra morphisms, and let $\langle s, t\rangle
\in B$. Given $X\in f(s)$, we know that $X =\Bild{\pi_{S}}{Z}$ for
some $Z\in h(s, t)$. Thus we find for any $t'\in Y$ some $s'\in X$
with $\langle s', t'\rangle \in B$. The symmetric property of a bisimulation
is established exactly in the same way. Hence $B$ is a bisimulation for
$(S, f)$ and $(T, g)$.
\EndProof
} 

We will now turn to
bisimulations for stochastic systems. A bisimulation relates two transition systems which are
connected through a mediating system. In order to define this for the
present context, we extend the crucial notion of morphisms
slightly in a  straightforward manner; this will be
helpful in the discussions to follow. 

\BeginDefinition{stoch-morphism-gen}
A \emph{morphism}\MMP{Morphism, again}
$m = (f, g): K_{1}\to K_{2}$ for 
stochastic relations
$K_{i}: (X_{i}, {\cal A}_{i})\Trans (Y_{i}, {\cal B}_{i})$
($i = 1, 2$) over general measurable spaces is given through the
measurable maps $f: X_{1}\to X_{2}$ and $g: Y_{1}\to Y_{2}$ such that
this diagram of measurable maps commutes
\begin{equation*}
\xymatrix{
(X_{1}, {\cal A}_{1})\ar[d]_{K_{1}}\ar[rr]^{f}&& (X_{2}, {\cal A}_{2})\ar[d]^{K_{2}}\\
\SubProb{Y_{1}, {\cal B}_{1}}\ar[rr]_{\SubProb{g}}&&\SubProb{Y_{2}, {\cal B}_{2}}
}
\end{equation*}
\EndDefinition
Equivalently, $K_{2}(f(x_{1})) = \SubProb{g}(K_{1}(x_{1}))$, which translates to $K_{2}(f(x_{1}))(B) = K_{1}(x_{1})(\InvBild{g}{B})$ for all $B\in{\cal B}_{2}$. 

\BeginDefinition{6-Bisimilar} 
The stochastic relations $K_{i}: (X_{i}, {\cal A}_{i})\Trans (Y_{i},
{\cal B}_{i})$ ($i = 1, 2$), are called
\emph{bisimilar}\MMP{Bisimilarity, $\SubProbSenza$} iff there exist a stochastic
relation $M: (A, {\cal X})\Trans (B, {\cal Y})$ and surjective
morphisms $m_{i} = (f_{i}, g_{i}): M\to K_{i}$ such that the $\sigma$-algebra
$ \InvBild{g_{1}}{{\cal B}_{1}} \cap \InvBild{g_{2}}{{\cal B}_{2}} $ is nontrivial,
i.e., contains not only $\emptyset$ and $B$.  The relation $M$ is
called \emph{mediating}.
\EndDefinition

The first condition on bisimilarity is in accordance with the general
definition of bisimilarity of coalgebras; it requests that $ m_{1} $
and $ m_{2} $ form a span of morphisms
\begin{equation*}
\xymatrix{
K_{1} & M\ar[l]_{m_{1}}\ar[r]^{m_{2}} & K_{2}.
}
\end{equation*}
Hence, the following diagram of measurable maps is supposed to commute with $m_{i} = (f_{i}, g_{i})$ for $i = 1, 2$
\begin{equation*}
\xymatrix{
(X_{1}, {\cal A}_{1})\ar[d]_{K_{1}} 
&& (A, {\cal X})\ar[ll]_{f_{1}}\ar[rr]^{f_{2}}\ar[d]^{M} 
&& (X_{2}, {\cal A}_{2})\ar[d]^{K_{2}}\\
\SubProb{Y_{1}, {\cal B}_{1}} 
&& \SubProb{B, {\cal Y}}\ar[ll]^{\SubProb{g_{1}}}\ar[rr]_{\SubProb{g_{2}}} 
&& \SubProb{Y_{2}, {\cal B}_{2}}
}
\end{equation*}
Thus, for each $a \in A, D \in{\cal B}_{1}, E \in {\cal B}_{2}$
the equalities 
\begin{alignat*}{2}
K_{1}\bigl(f_{1}(a)\bigr)(D) & = \bigl(\SubProb{g_{1}}\circ  M\bigr)(a)(D) && = M(a)\bigl(\InvBild{g_{1}}{D}\bigr)\\
K_{2}\bigl(f_{2}(a)\bigr)(E) & = \bigl(\SubProb{g_{2}}\circ  M\bigr)(a)(E) && = M(a)\bigl(\InvBild{g_{2}}{E}\bigr)
\end{alignat*}
should be satisfied.
The second condition, however, is special; it states that we can find an event $ C^* \in{\cal Y} $ which is common to both $K_{1}$ and $K_{2}$ in the sense that
\begin{equation*}
\InvBild{g_{1}}{B_{1}} = C^* = \InvBild{g_{2}}{B_{2}}
\end{equation*}
for some
$
B_{1} \in {\cal B}_{1}
$
and
$
B_{2} \in {\cal B}_{2}
$
such that both $C^* \not= \emptyset$ and $C^* \not= B$ hold (note that
for $C^* =\emptyset$ or $C^* = B$ we can always take the empty and the full set,
respectively). Given such a $C^*$ with $B_{1}, B_{2}$ from above we get for each $a \in
A$
\begin{equation*}
  K_{1}(f_{1}(a))(B_{1})
= 
  M(a)(\InvBild{g_{1}}{B_{1}}) 
=  
  M(a)(C^*)
=  
  M(a)(\InvBild{g_{2}}{B_{2}})
=  
  K_{2}(g_{2}(a))(B_{2});
\end{equation*}
thus the event $C^*$ ties $K_{1}$ and $K_{2}$ together. Loosely
speaking, $ \InvBild{g_{1}}{{\cal B}_{1}} \cap \InvBild{g_{2}}{{\cal
    B}_{2}} $ can be described as the $\sigma$-algebra of common
events, which is required to be nontrivial.  

Note that without the second condition two relations $K_{1}$ and
$K_{2}$  would always be bisimilar:
Put $ A := X_{1}\times X_{2}$, $B :=Y_{1}\times Y_{2} $ and set for
$\langle x_{1}, x_{2}\rangle \in A$ as the mediating relation $
M(x_{1}, x_{2}) := K_{1}(x_{1}) \otimes K_{2}(x_{2});$ that is, define
$M$ pointwise to be the product measure of $K_{1}$ and $K_{2}$. Then
the projections will make the diagram commutative. But although this
notion of bisimilarity is sometimes suggested, it is
way too weak, because bisimulations relate transition systems, and
it does not promise particularly interesting insights when two arbitrary systems can
be related. It is also clear that using products for mediation does not work for the
subprobabilistic case. But the definition above captures the general
case as well.

\section{Expressivity of Kripke Models}
\label{sec:kripke-models}

Transition kernels will be used now for interpreting modal logics. Consider this grammar for formulas
\begin{equation*}
  \phi ::= \top \mid \phi_{1}\wedge\phi_{2}\mid \Diamond_{q}\phi
\end{equation*}
with $q\in\Rational, q \geq 0$. Note that the logic is negation free
and has on the propositional level only conjunction;
this may be motivated by the observation that we work in a Boolean
set algebra in which negation is available.

The informal interpretation in a
probabilistic transition system is that $\top$ always holds, and that
$\Diamond_{q}\phi$ holds with probability not smaller than $q$ after a
transition in a state in which formula $\phi$ holds. Now let $K:
X\Trans X$ be a transition kernel for the measurable space $X$, and define inductively
\begin{align*}
  \Gilt[\top]_K :=&\ X\\
\Gilt[\phi_{1}\wedge\phi_{2}]_K :=&\ \Gilt[\phi_{1}]_K\cap\Gilt[\phi_{2}]_K\\
\Gilt[\Diamond_{q}\phi]_K :=&\ \{x\in X \mid K(x)(\Gilt_K)\geq q\}\\
=&\ \InvBild{K}{{\theBeta{\Gilt_K}{q}{{\cal A}}}}
\end{align*}
One shows by induction on the structure of the formula that
the sets $\Gilt_K$ are measurable, since $K$ is a stochastic
relation. We say that $\phi$ holds in $x\in X$ iff $x\in \Gilt_{K}$;
this is also written as $K, x\models \phi$. Note that $\phi$ does not
hold in $x$ iff $x\in X\setminus\Gilt_{K}$, the latter set being
measurable. This observation supports the decision to omit negation as
an operator.   

\medskip

One usually takes a set of actions and defines modalities
$\langle a\rangle_{r}$ for action $a$, generalizing
$\Diamond_{r}$. For the sake of argument, I will stick for the time
being to the very simple case of having only one action. The arguments
for the general case will be exactly the same, taking into account
that one deals with a family of stochastic relations rather than with
one relation. I will also do without primitive formulas and introduce
them only when I need them; they do not add to the argument's
substance right now. A \emph{Kripke model}\MMP{Kripke model} with
state space $X$ and transition law $K$ is just a
$\SubProbSenza$-coalgebra $(X, K)$, for the extensions see
page~\pageref{modifications}.

Define for state $x\in X$ its \emph{theory}\MMP{$\theTheory{K}{x}$} by
\begin{equation*}
  \theTheory{K}{x} := \{\phi\mid x\in\Gilt_{K}\} = \{\phi\mid K, x\models \phi\}
\end{equation*}

For comparing the expressivity\MMP{Expressivity} of Kripke models, we use these
approaches
\begin{itemize}
\item $(X, K)$ is \emph{logically equivalent} to $(Y, L)$ iff 
\begin{equation*}
\{\theTheory{K}{x}\mid x\in X\} = \{\theTheory{L}{y}\mid y\in Y\},
\end{equation*}
thus iff given a state $x\in X$, there exists a state $y\in Y$ with
exactly the same theory, and vice versa.
\item $(X, K)$ is \emph{behaviorally equivalent} to $(Y, L)$ iff
there exists a Kripke model $(Z, M)$ and surjective morphisms 
\begin{equation*}
\xymatrix{
(X, K)\ar[rr]^{f} && (Z, M) && (Y, L)\ar[ll]_{g}.
}
\end{equation*}
Model $(Z, M)$\MMP{Span, co-span} is called \emph{mediating} (and the diagram a \emph{co-span}).
\item $(X, K)$ is \emph{bisimilar} to $(Y, L)$ iff
there exists a Kripke model $(Z, M)$ and surjective morphisms 
\begin{equation*}
\xymatrix{
(X, K) && (Z, M)\ar[ll]_{f}\ar[rr]^{g} && (Y, L)
}
\end{equation*}
such that the $\sigma$-algebra of common events is not trivial. Model
$(Z, M)$ is also called \emph{mediating} (and the diagram a \emph{span}).
\end{itemize}

We will investigate these notions of expressivity now. Note that there
are many variants to bisimilarity, e.g., state bisimulations, but
life is difficult enough, so I will not not deal with them here. 

The key property in this business is

\BeginProposition{preserves}
Let $(X, K)$ and $(Y, L)$ be Kripke models and a morphism $f: (X,
K)\to (Y, L)$. Then 
\begin{equation*}
  K, x\models \phi \text{ iff }L, f(x)\models \phi
\end{equation*}
holds for all $x\in X$ and all formulas $\phi$. 
\EndProposition

\BeginProof
The assertion is equivalent to $\Gilt_{K} = \InvBild{f}{\Gilt_{L}}$
for all $\phi$. This is clear for $\phi=\top$, and if it is true for
formulas $\phi_{1}$ and $\phi_{2}$, then it is true also for
$\phi_{1}\wedge\phi_{2}$. The interesting case is
$\phi_{1}=\Diamond_{r}\phi$:
\begin{align*}
  \Gilt[\Diamond_{r}\phi]_{K} & = 
\{x\in X\mid K(x)(\Gilt_{K})\geq r\}&\text{ (definition)}\\
& = \{x\in X\mid K(x)(\InvBild{f}{\Gilt_{L}})\geq r\}&\text{
                                                       (induction hypothesis)}\\
& = \{x\in X\mid L(f(x))(\Gilt_{L}) \geq r\}& \text{ ($f$ is a morphism)}\\
& = \InvBild{f}{\Gilt[\Diamond_{r}\phi]_{L}}
\end{align*}
\EndProof

This provides us with an easy consequence:

\BeginCorollary{beh-impl-log}
Behaviorally equivalent Kripke models are logically equivalent, so are
bisimilar Kripke models. 
\EndCorollary

The classical Hennessy-Milner Theorem~\cite[Theorem
2.7.32]{EED-Companion} for transition systems states that logically
equivalent models are behaviorally equivalent, provided the models are
image finite. This condition used to prevent the system from becoming too
large, but is difficult to model for stochastic
Kripke models. Hence we need a condition which is intended restrict the size of
the system, so we need a condition for smallness. Here we proceed as
follows.

Fix a Kripke model $(X, K)$. The logic induces an equivalence
relation $\alpha$ on $X$ upon setting
\begin{equation*}
  \isEquiv{x}{x'}{\alpha} \Leftrightarrow \bigl[K, x\models \phi
  \text{ iff }K, x'\models \phi\bigr]\text{  for all formulas }\phi.
\end{equation*}
Thus $\isEquiv{x}{x'}{\alpha}$ iff the logic cannot distinguish
between $x$ and $x'$. The discussion above shows that we may factor
$X$, obtaining the factor space $\Faktor{X}{\alpha}$. Moreover we know
that $K(x)(\Gilt_{K}) = K(x')(\Gilt_{K})$, provided
$\isEquiv{x}{x'}{\alpha}$ (suppose there exists $r$ with
$K(x)(\Gilt_{K}) < r \leq K(x')(\Gilt_{K})$, hence $K, x'\models
\Diamond_{r}\phi$, but $K, x\not\models \Diamond_{r}\phi$). This means
that $\alpha$ is a congruence for $K$, the $\sigma$-algebra on the
factor space being generated by the $\alpha$-invariant sets
$\Sigma_{X, \alpha}$. On the other hand, we have the $\sigma$-algebra
$\Theta_{X} := \sigma(\{\Gilt_{K}\mid \phi\text{ is a
  formula}\})$ which is generated by the validity sets for the
formulas (recall that each validity set is measurable). Since each
validity set is $\alpha$-invariant, we have $\Theta_{X}\subseteq
\Sigma_{X, \alpha}$. It may happen, however, that the containment is
proper~\cite[Example 2.6.7]{EED-CoalgLogic-Book}.

\BeginDefinition{Kripke-small}
The Kripke model is said to be \emph{small}\MMP{Small Kripke model} iff $\Sigma_{\alpha} =
\sigma\bigl(\{\Gilt_{K}\mid \phi \text{ is a formula}\}\bigr)$.
\EndDefinition

Smallness will permit us to establish an analogon to the
Hennessy-Milner Theorem; technically, it says that the $\sigma$-algebra
on the factor space is determined by the images of the validity sets
under the factor map $\rho_{\alpha}$:

\BeginLemma{lemma-small}
If $(X, K)$ is small, then $\Faktor{\mathcal{A}}{\alpha} =
\sigma\bigl(\{\Bild{\rho_{\alpha}}{\Gilt_{K}}\mid \phi\text{ is a
  formula}\}\bigr)$. 
\QED
\EndLemma

Smallness also ensures that we have a $\cap$-stable generator for the
factor $\sigma$-algebra (note that $\Bild{\rho_{\alpha}}{A\cap B}
=\Bild{\rho_{\alpha}}{A}\cap\Bild{\rho_{\alpha}}{B}$ if $A$ and $B$
are $\alpha$-invariant) , so that we have ---~in view of the
$\pi$-$\lambda$-Theorem~--- a fairly practical handle to
deal with the factor space. This will be seen in a moment.

Recall that topological space is called \emph{Polish}\MMP{Polish, analytic} iff it is second
countable, and its topology can be metrized by a complete metric;
examples include the reals $\Real$, $[0, 1]^{\Nat}$, the bounded continuous
functions over $\Real$, and $\SubProbSenza{X}$, if $X$ is Polish; the
rationals $\Rational$ are \textbf{not} Polish. A
measurable image of a Polish space is called an \emph{analytic} space. Polish
spaces, and to some extent, analytic spaces, have very convenient
measure theoretic properties (for a general and accessible account,
see~\cite{Srivastava}, for a discussion tailored towards the purposes
discussed here, see~\cite[Sections 4.3, 4.4]{EED-Companion}).

We note that a Kripke model over a Polish space is always small. This is so because the
equivalence relation induced by the logic is \emph{smooth}\MMP{Smooth}, i.e., countably
generated, because the logic has only countably many formulas. The
observation follows then from some general properties about smooth
equivalence relations on Polish spaces~\cite[Proposition
4.4.26]{EED-Companion}.

Now let $(X, K)$ and $(Y, L)$ be small models which are logically
equivalent. We want to show that they are behaviorally equivalent, so
we have to construct a mediator. Let $\alpha_{K}$ resp. $\alpha_{L}$
be the corresponding equivalence relations with classes
$\Klasse{\cdot}{K}$ and $\rho_{K}$ as factor map, similarly for $L$. Define
\begin{align*}
  \Re &:= \{\langle x, y\rangle\mid  \theTheory{K}{x} =
        \theTheory{L}{y}\},\\
\Re_{0} & := \{\langle\Klasse{x}{K}, \Klasse{y}{L}\rangle\mid \langle
          x, y\rangle\in\Re\}
\end{align*}
Since the models are logically equivalent, $\Re$ is both left and
right total. We will show now that $\Re_{0}$ is the graph of a
bi-measurable map $f: \Faktor{X}{R}\to \Faktor{Y}{L}$. From the
construction of $\Re_{0}$ it is clear that $f$ is a bijection, so we
have to cater for measurability. This is done through the principle of
good sets in conjunction with the $\pi$-$\lambda$-Theorem. Consider 
\begin{equation*}
  \mathcal{D} := \{B\subseteq\Faktor{Y}{L}\text{ measurable}\mid
  \InvBild{f}{B}\subseteq\Faktor{X}{K}\text{ is measurable}\}.
\end{equation*}
Then $\mathcal{D}$ is closed under complementation and under countable
disjoint unions. Let $\phi$ be a formula, then 
\begin{equation*}
  \InvBild{f}{\Bild{\rho_{L}}{\Gilt_{L}}} = \Bild{\rho_{K}}{\Gilt_{K}}.
\end{equation*}
Thus $\Bild{\rho_{L}}{\Gilt_{L}}\in\mathcal{D}$ for each formula
$\phi$. Since $\{\Bild{\rho_{L}}{\Gilt_{L}}\mid \phi\text{ is a
  formula}\}$ is a $\cap$-stable generator of the factor
$\sigma$-algebra due to $(Y, L)$ being small, we conclude from the $\pi$-$\lambda$-Theorem that
$\mathcal{D}$ equals the $\sigma$-algebra of all measurable sets of
$\Faktor{Y}{L}$. This shows that $f$ is measurable, the argumentation
is exactly the same for $f^{-1}$. Consequently, $\Re_{0}$ is the graph
of an isomorphism.

Now look at this diagram:
\begin{equation*}
\xymatrix{
X\ar[d]_{\rho_{K}} && Y\ar[d]^{\rho_{L}}\\
\Faktor{X}{\alpha_K}\ar@{=}[rr]&&\Faktor{Y}{\alpha_L}
}
\end{equation*}
Since $\rho_{K}$ as well as $\rho_{L}$ are morphisms $(X, K)\to
(\Faktor{X}{\alpha_{K}}, K_{\alpha})$ resp. $(Y, L)\to
(\Faktor{Y}{\alpha_{L}}, L_{\alpha})$, we have established this
counterpart to the Hennessy-Milner Theorem:

\BeginProposition{small-log-beh}
Small logically equivalent Kripke models are behaviorally
equivalent. \QED
\EndProposition

What about bisimilarity? Rutten's paper~\cite{Rutten} gives a calculus
of bisimilarity for coalgebras. Unfortunately the really interesting properties
assume that the functor under consideration preserves
weak pullbacks. This is not the case for the Giry functor, as the
following example demonstrates. It presents a situation in which no semi-pullback
exists. A first example in this direction was suggested
in~\cite[Theorem 12]{Sanchez}. It is based on the extension of
Lebesgue measure to a $\sigma$-algebra which does contain the Borel
sets of $[0, 1]$ augmented by a non-measurable set, and it shows that
one can construct Markov transition systems which do not have a
semi-pullback. The example below simplifies this by showing that one does not have to
consider transition systems, but that a look at the measures on which
they are based suffices.

{
\def\FinM{\SubProb}
\BeginExample{no-semi-pullback-exists}
A morphism $f: (X, {\cal A}, \mu)\to (Y, {\cal B}, \nu)$ of measure
spaces is an ${\cal A}$-${\cal B}$-measurable map $f: X\to Y$ such
that $\nu = \FinM{f}(\mu)$. Since each finite measure can be viewed as
a transition kernel, this is a special case of morphisms for
transition kernels. If ${\cal B}$ is a sub-$\sigma$-algebra of
${\cal A}$ with $\mu$ an extension to $\nu$, then the identity is a
morphisms $(X, {\cal A}, \mu)\to (X, {\cal B}, \nu)$.

Denote Lebesgue measure on $([0, 1], \Borel{[0, 1]})$ by
$\lambda$. Assuming the Axiom of Choice, we know that there exists
$W\subseteq [0, 1]$ with $\lambda_{*}(W) = 0$ and
$\lambda^{*}(W) = 1$. Here $\lambda_{*}$ and $\lambda^{*}$ denote the
inner resp. outer measure associated with Lebesgue measure. The
technical condition says that whenever we have a measurable set
$P\subseteq W$, then $\lambda(P) = 0$ must hold, and whenever we have
a measurable set $Q$ with $W\subseteq Q$, then $\lambda(Q) = 1$. These
conditions make sure that $W$ is not in the universal completion of
$[0, 1]$, which means that Lebesgue measure cannot be extended uniquely to it
in a canonic way. 

But we find other, less canonic extensions, actually, plenty of them. Denote by
\begin{equation*}
{\cal A}_{W} := \sigma(\Borel{[0, 1]}\cup\{W\})
\end{equation*}
the smallest
$\sigma$-algebra containing the Borel sets of $[0, 1]$ and $W$. We 
find for each $\alpha\in[0, 1]$ a measure $\mu_{\alpha}$ on
${\cal A}_{W}$ which extends $\lambda$ such that
$\mu_{\alpha}(W) = \alpha$ by~\cite[Exercise 4.6]{EED-Companion}.

Hence by the remark just made, the identity yields a morphism
\begin{equation*}
f_{\alpha}: \bigl([0, 1], {\cal A}_{W}, \mu_{\alpha}\bigr) \to ([0, 1],
\Borel{[0, 1]}, \lambda).
\end{equation*}
Now let $\alpha\not=\beta$, then
\begin{equation*}
\xymatrix{
\bigl([0, 1], {\cal A}_{W}, \mu_{\alpha}\bigr)\ar[rr]^{f_{\alpha}} &&
 \bigl([0, 1], \Borel{[0, 1]}, \lambda\bigr) &&
 \bigl([0, 1], {\cal A}_{W}, \mu_{\beta}\bigr)\ar[ll]_{f_{\beta}}
}
\end{equation*}
is a co-span of morphisms.  We claim that this co-span does not have a
semi-pullpack. In fact, assume that $(P, {\cal P}, \rho)$ with
morphisms $\pi_{\alpha}$ and $\pi_{\beta}$ is a semi-pullback, then
$f_{\alpha}\circ \pi_{\alpha} = f_{\beta}\circ \pi_{\beta}$, so that
$\pi_{\alpha} = \pi_{\beta}$, and
$\InvBild{\pi_{\alpha}}{W} = \InvBild{\pi_{\beta}}{W}\in{\cal P}$. But
then
\begin{equation*}
  \alpha = \mu_{\alpha}(W) = \rho(\InvBild{\pi_{\alpha}}{W}) = \rho(\InvBild{\pi_{\beta}}{W}) = \mu_{\beta}(W) = \beta.
\end{equation*}
This contradicts the assumption that $\alpha\not=\beta$. 
\EndExample
}

The question whether behaviorally equivalent Kripke models are
bisimilar was open for quite some time, until Desharnais, Edalat and
Panangaden showed in \cite{Desharnais-Edalat-Panangaden,Edalat} that
behaviorally equivalent Kripke models based on analytic spaces are
bisimilar with an analytic mediating system. This result was sharpened
in~\cite{EED-Analytic,EED-SemiPullback}: if the contributing models are
based on Polish spaces, there exists a mediator based on a Polish
space as well. Interestingly, the proof techniques are very
different. While the analytic case is delt with using conditional
expectations, which are known to exist in analytic spaces, the Polish
case is based on a selection argument, which ---~strange enough~--- does not
generalize to analytic spaces.

The following proposition summarizes the discussion (see~\cite{Desharnais-Edalat-Panangaden}
for the analytic case, and~\cite[Proposition 4.10.20]{EED-Companion} for
the Polish case):

\BeginProposition{6-HasSemiPullbacks} 
Let $(X_{i}, K_{i})$ be Kripke models over analytic spaces $X_{1}$,
$X_{2}$, and assume that $(X, K)$ is a stochastic relation, where $X$
is a second countable metric space. Assume that we have a co-span of
morphisms $m_{i}: K_{i}\to K, i = 1, 2$, then there exists a
stochastic relation $M$ and morphisms
$m^{+}_{i}: M\Trans K_{i}, i = 1, 2$ rendering this diagram
commutative. 
\begin{equation*}
\xymatrix{
M\ar[rr]^{m^{+}_1}\ar[d]_{m^{+}_2}&&K_2\ar[d]^{m_2}\\
K_1\ar[rr]_{m_1}&&K
}
\end{equation*}
The stochastic relation $M$ is defined over an analytic space. If
$X_{1}$ and $X_{2}$ are Polish, $M$ can be defined over a Polish
space. 
\QED
\EndProposition 

Proposition~\ref{6-HasSemiPullbacks} is the crucial step in
establishing~\cite[Proposition 4.10.22]{EED-Companion}:

\BeginProposition{log-bisim}
Logically equivalent Kripke models over analytic spaces are
bisimilar. The mediating model is analytic again. If the contributing
models are Polish, the mediator is Polish as well.
\QED
\EndProposition

These results\MMP{Modification 1} are formulated for Kripke models for the basic modal
language with the grammar
\begin{equation*}
  \phi ::= \top \mid \phi_{1}\wedge\phi_{2}\mid \Diamond_{r}\phi
\end{equation*}
with $r\in [0, 1]\cap\Rational$.\label{modifications} They generalize easily to a modal
logic in which the modalities are given through $\langle a\rangle_{r}$
for some action $a\in A$. Here we associate with each action $a$ a
stochastic relation $K_{a}: X \Trans X$. A \emph{Kripke model} is then given
through $\bigl(X, (K_{a})_{a\in A}\bigr)$, and a morphism $f: \bigl(X,
(K_{a})_{a\in A}\bigr)\to \bigl(Y, (L_{a})_{a\in A}\bigr)$ is then a
measurable map $f: X\to Y$ such that $L_{a}\circ f = \SubProb{f}\circ
K_{a}$ holds for all $a\in A$. 

Another modification\MMP{Modification 2}
addresses the introduction of primitive formulas. The formulas for the
general modal
logic now look as follows:
\begin{equation*}
  \phi ::= p \mid \top \mid \phi_{1}\wedge\phi_{2}\mid \langle a\rangle_{r}\phi
\end{equation*}
with $p\in \AE$ a primitive formula, $a\in A$ an action, and $r\in [0,
1]\cap\Rational$. A Kripke model is now given through 
$\bigl(X, (V_{p})_{p\in P}, (K_{a})_{a\in A}\bigr)$, where
$V_{p}\subseteq X$ is a measurable subset for each $p\in \AE$,
indicating the set of states in which a primitive formula holds;
accordingly, we put $\Gilt[p] := V_{p}$. A
morphism $f: \bigl(X, (V_{p})_{p\in \AE}, (K_{a})_{a\in A}\bigr)\to
\bigl(Y, (W_{p})_{p\in \AE}, (L_{a})_{a\in A}\bigr)$ is defined as above
with the additional requirement that 
$
  \InvBild{f}{W_{p}} = V_{p}
$
holds for each $p\in \AE$. 

\medskip

The grammar above will be the one to use in the sequel as a sort of
shell, where we fill in specific sets of actions. 

\section{Stochastic Effectivity Functions}
\label{sec:stoch-effect-funct}

We will now look into the interpretation of game logics; the reader
interested in computational aspects is referred to~\cite{Parikh-Games1985}. In terms of
modal logics, the actions are games, so the modalities are not flat,
but rather structured according to the grammar through which games are
specified. We will first make some general remarks.

Angel and Demon play against each other, taking turns. The two person
game is modelled by this grammar\MMP{Two person game}
\begin{equation*}
\tau ::= \gamma~\mid~\tau^d~\mid~\tau_1\cup\tau_2~\mid~\tau_1\cap\tau_2~\mid~\tau_1;\tau_2~\mid~\tau^*~\mid~\tau^\times
\end{equation*}
with $\gamma\in\Gamma$, the set of atomic games. 
Games can be combined in different ways. If $\tau$ and $\tau'$ are
games, $\tau;\tau'$ is the sequential composition of $\tau_1$ and
$\tau_2$, so that plays $\tau$ first, then $\tau'$. In the game
$\tau\cup\tau'$, Angel has the first move and decides whether $\tau$
or $\tau'$ is to be played, then the chosen game is played;
$\tau\cup\tau'$ is called the \emph{angelic choice} between $\tau_1$
and $\tau_2$. Similarly, in $\tau\cap\tau'$ Demon has the first move
and decides whether $\tau$ or $\tau'$ is to be played; accordingly,
$\tau\cap\tau'$ is the \emph{demonic choice} between the games. In the
game $\tau^*$, game $\tau$ is played repeatedly, until Angel decides
to stop; it is not said in advance how many times the game is to be
played, but it has to stop at some time; this is called \emph{angelic
  iteration}. Dually, Demon decides to stop for the game
$\tau^\times$; this is called \emph{demonic iteration}. Finally, the
rôles of Angel and Demon are interchanged in the game $\tau^d$, so all
decisions made by Demon are now being made by Angel, and vice versa. 

When writing down games, we assume for simplicity that composition
binds tighter than angelic or demonic choice. We make these
assumptions~\cite{Pauly-Parikh,Parikh-Games1985}:
\begin{dingautolist}{192}
\item\label{anfang} $(\tau^d)^d$ is identical to $\tau$ (recall that $\cdot^{d}$
  indicates Angel and Demon switching rôles). 
\item Demonic choice can be represented through angelic choice: The game $\tau_1\cap\tau_2$ coincides with the game $(\tau_1^d\cup\tau_2^d)^d$.
\item Similarly, demonic iteration can be represented through its angelic counterpart: $(\tau^\times)^d$ is equal to $(\tau^d)^*$,
\item Composition is right distributive with respect to angelic choice: Making a decision to play $\tau_1$ or $\tau_2$ and then playing $\tau$ should be the same as deciding to play $\tau_1;\tau$ or $\tau_2;\tau$, thus $(\tau_1\cup\tau_2);\tau$ equals $\tau_1;\tau\cup\tau_2;\tau$. 

Note that left distributivity would mean that a choice between
$\tau;\tau_1$ and $\tau;\tau_2$ is the same as playing first $\tau$
then $\tau_1\cup\tau_2$; this is a somewhat restrictive assumption,
since the choice of playing $\tau_1$ or $\tau_2$ may be a decision
made by Angel only after $\tau$ is
completed~\cite[p. 191]{Benthem-LogicGames}. Thus we do not assume
this in general (it can be shown, however, that in Kripke generated
models these choices are in fact equivalent~\cite[Proposition 4.9.40]{EED-Companion}).
\item\label{star-assumption}
We assume similarly that 
$
\tau^*;\tau_0 
\text{ equals }
\tau_0\cup\tau^*;\tau;\tau_0.
$
Hence when playing $\tau^*;\tau_0$  Angel may decide to play $\tau$ not at all and to continue with $\tau_0$ right away, or to play $\tau^*$ followed by $\tau;\tau_0$. Thus $\tau^*;\tau_0$ expands to 
$
\tau_0 \cup \tau;\tau_0 \cup \tau;\tau;\tau_0 \cup \dots.
$
\item $(\tau_1;\tau_2)^d$ is the same as $\tau_1^d;\tau_2^d$.
\item\label{ende} Angelic and demonic choice are commutative and associative,
  composition is associative. 
\end{dingautolist}

For arriving at an interpretation, some historic remarks are helpful,
and in order. Parikh~\cite{Parikh-Games1985}, and later
Pauly~\cite{Pauly-CWI} propose interpreting game logic through a
neighborhood model. Assign to each primitive game $\gamma$ and each player
(Angel: $A$; Demon: $D$) a neighborhood relation
$N_g^{(i)} \subseteq S \times \PowerSet{S}\ (i \in\{A, D\})$\MMP{$N_g^{(i)}$} with the
understanding that $x N_{\gamma}^{(i)} S$ indicates player $i$ having a
strategy in state $x$ to force a state in $S \subseteq X$. Here $X$ is
the set of states over which the game is interpreted. The fact that
$x N_{\gamma}^{(i)} S$ is sometimes described by saying that player $i$ is
effective for $S$ (with game $\gamma$ in state $x$). It is desirable that
$x N_g^{(i)} S$ and $S \subseteq S'$ imply $x N_g^{(i)} S'$ for all
states $x$. We assume that the game is \emph{determined}, i.e., that
exactly one of the players has a winning strategy\MMP{Determined game}. Thus
$S \subseteq X$ is effective for player $A$ in state $x$ if and only
if $X \setminus S$ is not effective for player $D$ in that
state. Consequently,
\begin{equation*}
x N_{\gamma}^{(D)} S \Leftrightarrow\neg(x N_g^{(A)} X\setminus S), 
\end{equation*}
which in turn implies that we only have to cater for Angel. We will
omit the superscript from the neighborhood relation
$N_{\gamma}$. Define the map
$ H_{\gamma}: X \to \PowerSet{\PowerSet{X}} $ upon setting
$$ H_{\gamma}(x) := \{S \subseteq X \mid x N_{\gamma} S\}, $$ then
$H_{\gamma}(x)$ is for all $x \in X$ an upper closed subset of
$\PowerSet{X}$ from which relation $N_{\gamma}$ can be recovered. This
function is called the \emph{effectivity function} associated with
relation $N_{\gamma}$. From $N_{\gamma}$ another map
$ \widetilde{N}_{\gamma}: \PowerSet{X}\to\PowerSet{X} $ is obtained
upon setting
$$ \widetilde{N}_{\gamma}(A) := \{x \in X \mid x N_{\gamma} A\} = \{x
\in X \mid A \in H_{\gamma}(x)\}.  $$
Thus state $x$ is an element of $\widetilde{N}_{\gamma}(A)$ iff Angel
has a strategy to force the outcome $A$ when playing $\gamma$ in $x$;
$\widetilde{N}_{\gamma}$ is actually a natural transformation. The
operations on games can be taken care of through this family of maps,
e.g., one sets recursively
\begin{align*}
\widetilde{N}_{\tau_1\cup \tau_2}(A) & := \widetilde{N}_{\tau_1}(A)\cup \widetilde{N}_{\tau_2}(A),\\
\widetilde{N}_{\tau_1;\tau_2}(A) & := (\widetilde{N}_{\tau_1}\circ \widetilde{N}_{\tau_2})(A),\\
\widetilde{N}_{\tau^*} & := \bigcup_{n \geq 0} \widetilde{N}_{\tau^n}(A).
\end{align*}
This refers only to Angel, Demon
is accommodated through
$A \mapsto S\setminus N_{\gamma}(X \setminus A)$ for primitive game
$\gamma$, and by the rules~\ref{anfang}~to~\ref{ende} from above. The maps $\widetilde{N}_{\tau}$
serve in Parikh's original paper as a basis for defining the
semantics of game logic. They are in one-to-one correspondence with
effectivity functions, hence effectivity functions are the main
actors. 

For a probabilistic interpretation of game logic, it turns out to be
convenient to also use effectivity functions as maps to upper closed
subsets. \textbf{But subsets of what?}

We observe these requirements for the \emph{portfolio}, i.e., for the sets
comprising the effectivity function\MMP{Portfolio, requirements}:
\begin{enumerate}
\item The elements of the sets should be probability measures. This is
  so because we want to force a distribution over the states, rather
  than a state proper.
\item The portfolio should consist of measurable sets, so that we can
  measure them.
\item Stochastic relations should be a special case, hence it should
  be possible to integrate them swiftly. 
\end{enumerate}

%
%
\newcommand{\bas}[2]{\ensuremath{\beta(#1, #2)}}
\def\VauSenza{\ensuremath{\dasF}}

Hence we require measurable sets of
probabilities as possible outcomes, but this is not enough. We will also impose a condition on
measurability on the interplay between distributions on states and
reals for measuring the probabilities of sets of states. This will lead to
the definition of a stochastic effectivity function. 

Denote for a measurable space $X$ the
*-$\sigma$-algebra on $\SubProb{X}$ by $\Borel{\SubProb{X}}$, and put
 \begin{equation*}
\Vau{X} := \{V \subseteq \Borel{\SubProb{X}} \mid V \text{ is upper closed}\}.
\end{equation*} 
A measurable map\MMP{$\Vau{X},
  \Vau{f}$} $f: X \to Y$ induces a map $\Vau{f}: \Vau{X}\to\Vau{Y}$ upon setting
\begin{equation*}
\Vau{f}(V) := \{W \in \Borel{\SubProb{Y}} \mid \InvBild{\SubProb{f}}{W}\in V\}
\end{equation*}
for $V \in \Vau{X}$, then clearly $\Vau{f}(V)\in\Vau{Y}$.  

\medskip 

Note that $\Vau{X}$ has not been equipped with a $\sigma$-algebra, so
the usual notion of measurability between measurable spaces cannot be
applied. In particular, $\VauSenza$ is not an endofunctor on the
category of measurable spaces. We will not discuss functorial aspects
of $\VauSenza$ here, referring the reader
to~\cite{EED-Alg-Prop_effFncts} instead. 

It would be most convenient if we could work in a monad --- after all,
the semantics pertaining to composition of games is modelled
appropriately using a composition operator, as demonstrated through
the definition of $\widetilde{N}_{\tau_1;\tau_2}$ above. Markov
transition systems are based on the Kleisli morphisms for the Giry
monad, and the functor assigning each set upper closed subsets of the
power set form a monad as well, see page~\pageref{proud}. So one might want to capitalize on the
composition of these monads. Alas, it is well known that the
composition of two monads is not necessarily a monad, so this approach
does not work, and one has to resort to ad-hoc methods simulating the
properties of a monad (or of a Kleisli tripel).

Preparing for this, we require some properties pertaining to
measurability, when dealing with the composition of distributions when
discussing composite games. This will be provided in the following
way. Let $H$ be a measurable
subset of $\SubProb{X}\times[0, 1]$ indicating a quantitative
assessment of subprobabilities (a typical example could be
$$ 
\{\langle \mu, q\rangle \mid \mu \in \bas{A}{q}, 0 \leq q \leq 1\}
$$
for some measurable $A \subseteq X$)\MMP{Quantitative aspect}. Fix some real $q$ and consider the set
$$H_q := \{\mu \mid \langle \mu, q\rangle \in H\}$$ of all measures
evaluated through $q$. We ask for all states $s$ such that this set is
effective for $s$. They should come from a measurable subset of
$X$. It turns out that this is not enough, we also require the real
components being captured through a measurable set as well --- after
all, the real component will be used to be averaged, i.e., integrated,
over later on, so it should behave decently. 

This idea is captured in the following definition.

\BeginDefinition{t-measurability}
Call a map $P: X \to \Vau{X}$ \emph{t-measurable} iff 
$
\{\langle s, q\rangle \mid H_q\in P(s)\} \subseteq X\times[0, 1]
$
is measurable whenever 
$ 
H \subseteq \SubProb{X} \times [0, 1] 
$
is measurable.  
A \emph{stochastic effectivity function}\MMP{Stochastic effectivity function} $P$ on a measurable space $X$
is a t-measurable map $P: X\to\Vau{X}$.
\EndDefinition

Each stochastic relation gives rise to an effectivity function; this
is indicated in the next example. The converse question, viz., under
what conditions a stochastic effectivity function is generated by a
stochastic relation, is more interesting, but a bit more cumbersome to
answer. For the sake of completeness we indicate a characterization in
the Appendix, see Section~\ref{sec:eff-vs-stroch-rel}.

\BeginExample{stoch-rel-eff-fnct}
Let $K: X\Trans X$ be a stochastic relation, then 
\begin{equation*}
  P_{K}(s) := \{A\subseteq \SubProb{X}\text{ measurable}\mid  K(s)\in A\}
\end{equation*}
is a stochastic effectivity function.
\EndExample

The next example is a little more sophisticated. It converts a finite
transition system over a finite state space into an effectivity
function.

\BeginExample{trans-to-eff-fnct}
Let $X := \{1, \dots, n\}$ for some $n\in\Nat$, and take the power set as a $\sigma$-algebra. Then $\SubProb{X}$ can be identified with the compact convex set
\begin{equation*}\textstyle
  \Pi_{n} := \{\langle x_{1}, \dots, x_{n}\rangle\mid x_{i}\geq 0
  \text{ for }1\leq i \leq n, \sum_{i=1}^{n} x_{i} = 1\}.
\end{equation*}
Geometrically, $\Pi_{n}$ is the convex hull of the unit vectors $e_{i}$, $1\leq i
\leq n$; here $e_{i}(i) = 1$, and $e_{i}(j) = 0$
if $i\not= j$ is the $i$-th $n$-dimensional unit vector. The weak-*-$\sigma$-algebra is the
Borel-$\sigma$-algebra $\Borel{\Pi_{n}}$ for the Euclidean topology on
$\Pi_{n}$. 

Assume we have a transition system $\to_{X}$ on $X$, hence
a relation $\to_{X}\subseteq X\times X$. Put $succ(s) := \{s'\in X\mid
s\to_{X} s'\}$ as the set of a successor states for state $s$. Define for
$s\in X$ the set of weighted successors
\begin{equation*}\textstyle
  \kappa(s):= \{\sum_{s'\in succ(s)}\alpha_{s'}\cdot e_{s'}\mid
  \Rational\ni\alpha_{s'}\geq 0 \text{ for } s'\in succ(s),
  \sum_{s'\in succ(s)}\alpha_{s'} = 1\}
\end{equation*}
and the upper closed set
\begin{equation*}\textstyle
P(s)  := \{A\in\Borel{\Pi_{n}}\mid \kappa(s)\subseteq A\}
\end{equation*}
A set $A$ is in the portfolio for $P$ in state $s$ if $A$  contains all
rational distributions on the successor states. Here we restrict our
attention to these rational distributions, which are positive convex
combinations of the unit vectors with rational coefficients. 

Then $P$ can be shown to be a stochastic effectivity function on
$X$~\cite[Example 4.1.14]{EED-Companion}.
Actually, I don't know what happens when we admit real coefficients
(things may become very complicated, then, since measurability might get lost). 
\EndExample

\subsection{Effectivity Functions vs. Stochastic Relations}
\label{sec:eff-vs-stroch-rel}
The tools for investigating the converse to
Example~\ref{stoch-rel-eff-fnct} come from the investigation of
deduction systems for probabilistic logics. In fact, we are given a
set of portfolios and want to know under which conditions this set is
generated from a single subprobability. The situation is roughly
similar to the one observed with deduction systems, where a set of
formulas is given, and one wants to know whether this set can be
constructed as valid under a suitable model. Because of the
similarity, we may take some inspiration from the work on deduction
systems, adapting the approach proposed by
R. Goldblatt~\cite{Goldblatt-Deduction}. Goldblatt works with formulas
while we are interested foremost in families of sets; this permits a
technically somewhat lighter approach in the present scenario.

Let $S$ be a measurable space of states; we will deal with the
measurable sets $\mathcal{A}$ of $S$ explicitly, so they are no longer
swept under the carpet. 
 
We  first have a look at a relation $R \subseteq [0, 1] \times \mathcal{A}$ which models bounding probabilities from below. Intuitively, $\langle r, A\rangle\in R$ is intended to characterize the set $\bas{A}{\geq r}$.

\BeginDefinition{R-binding}
$R \subseteq [0, 1] \times \mathcal{A}$ is called a \emph{characteristic relation on $S$} iff these conditions are satisfied
\begin{align*}
\text{\ding{172}}\ &\frac{\langle r, A\rangle\in R, A \subseteq B}{\langle r, B\rangle\in R} &
\text{\ding{173}}\  &\frac{\langle r, A\rangle\in R, r \geq s}{\langle s, A\rangle\in R}\\
\text{\ding{174}}\  &\frac{\langle r, A\rangle\notin R, \langle s, B\rangle\notin R, r + s \leq 1}{\langle r+s, A\cup B\rangle\notin R}&
\text{\ding{175}}\   &\frac{\langle r, A\cup B\rangle\in R, \langle s,A \cup (S\setminus B)\rangle\in R, r + s \leq 1}{\langle r+s, A\rangle\in R}\\
\text{\ding{176}}\   &\frac{\langle r, A\rangle\in R, r + s > 1}{\langle s, S \setminus A\rangle\notin R}&
\text{\ding{177}}\   &\frac{\langle r, \emptyset\rangle\in R}{r = 0}\\
\text{\ding{178}}\  &\frac{A_1 \supseteq A_2 \supseteq \dots, \forall
                      n \in \Nat: \langle r, A_n\rangle\in R}{\langle
                      r, \bigcap_{n\geq 1} A_n\rangle \in R} &
                                                               \text{\ding{179}}&\langle
                                                               1,
                                                               S\rangle\in R
\end{align*}
\EndDefinition
The conditions $\text{\ding{172}}$ and $\text{\ding{173}}$ make sure that bounding from below is monotone both in its numeric and in its set valued component. By $\text{\ding{174}}$ and $\text{\ding{175}}$ we cater for sub- and superadditivity of the characteristic relation, condition $\text{\ding{177}}$ sees to the fact that the probability for the impossible event cannot be bounded from below but through $0$, and finally $\text{\ding{178}}$ makes sure that if the members of a decreasing sequence of sets are uniformly bounded below, then so is its intersection. These conditions are adapted from the S-axioms for T-deduction systems in~\cite[Section 4]{Goldblatt-Deduction}. An exception is \ding{178} which is weaker than the Countable Additivity Rule in~\cite[Definition 4.4]{Goldblatt-Deduction}; we do not need a rule as strong as the latter one because we work with sets, hence we can deal with descending chains of sets directly. 

We show that each characteristic relation defines a probability
measure; the proof follows \emph{mutatis mutandis}~\cite[Theorem
5.4]{Goldblatt-Deduction}.

\BeginProposition{def-measure}
Let $R \subseteq [0, 1] \times \mathcal{A}$ be a \emph{characteristic relation on $S$}, and define for $A\in\mathcal{A}$
\begin{equation*}
\mu_R(A) := \sup\{r \in [0, 1] \mid \langle r, A\rangle\in R\}.
\end{equation*} 
Then $\mu_R$ is a probability measure on $\mathcal{A}$. 
\QED
\EndProposition

\BeginProof
1. 
$\text{\ding{177}}$ implies that $\mu_R(\emptyset) = 0$, and $\mu_R$
is monotone because of $\text{\ding{172}}$. It is also clear that
$\mu_R(A) \leq 1$ always holds. We obtain from $\text{\ding{173}}$ that $\langle s,
A\rangle\notin R$, whenever $s \geq r$ with $\langle r, A\rangle\notin
R$. Trivially, $\text{\ding{179}}$ implies that $\mu_R(S) = 1$. 

2. 
Let $A_1, A_2\in\mathcal{A}$ be arbitrary. Then 
\begin{equation*}
\mu_R(A_1\cup A_2) \leq \mu_R(A_1) + \mu_R(A_2).
\end{equation*}
In fact, if 
$
\mu_R(A_1) + \mu_R(A_2) < q_1 + q_2 \leq \mu_R(A_1\cup A_2)
$
with 
$
\mu_R(A_i) < q_i\ (i = 1, 2),
$
then 
$
\langle q_i, A_i\rangle\notin R 
$
for $i = 1, 2$. Because $q_1 + q_2 \leq 1$, we obtain from $\text{\ding{174}}$ that 
$
\langle q_1 + q_2, A_1 \cup A_2\rangle\notin R.
$
By $\text{\ding{173}}$ this yields 
$
\mu_R(A_1\cup A_2) < q_1 + q_2,
$
contradicting the assumption. 

3. 
If $A_1$ and $A_2$ are disjoint, we observe first that 
$
\mu_R(A_1) + \mu_R(A_2) \leq 1.
$
Assume otherwise that we can find $q_i \leq \mu_R(A_i)$ for $i = 1, 2$ with $q_1 + q_2 > 1$. Because $\langle q_1, A_1\rangle\in R$ we conclude from $\text{\ding{176}}$ that 
$
\langle q_2, S\setminus A_1\rangle\notin R,
$
hence $\langle q_2, A_2\rangle\notin R$ by $\text{\ding{172}}$, contradicting $q_2 \leq \mu_R(A_2)$.  

This implies that 
\begin{equation*}
\mu_R(A_1) + \mu_R(A_2) \leq \mu_R(A_1\cup A_2).
\end{equation*}
Assuming this to be false, we find $q_1 \leq \mu_R(A_1), q_2 \leq \mu_R(A_2)$ with
\begin{equation*}
\mu_R(A_1 \cup A_2) < q_1 + q_2 \leq \mu_R(A_1) + \mu_R(A_2).
\end{equation*}
Because $\langle q_1, A_1\rangle\in R$, we find 
$
\langle q_1, (A_1\cup A_2)\cap A_1\rangle\in R, 
$
because $\langle q_2, A_2\rangle\in R$ we see that
$
\langle q_2, (A_1\cup A_2)\cap (S\setminus A_1)\rangle\in R 
$
(note that $(A_1\cup A_2)\cap A_1 = A_1$ and $(A_1\cup A_2)\cap (S\setminus A_1) = A_2$, since $A_1\cap A_2 = \emptyset$). From $\text{\ding{175}}$ we infer that
$
\langle q_1 + q_2, A_1 \cup A_2\rangle\in R,
$
so that 
$
q_1 + q_2 \leq \mu_R(A_1 \cup A_2),
$
which is a contradiction. 

Thus we have shown that $\mu_R$ is additive.

4.
From $\text{\ding{178}}$ it is obvious that 
\begin{equation*}
\mu_R(A) = \inf_{n\in\Nat} \mu_R(A_n),
\end{equation*}
whenever 
$
A = \bigcap_{n\in\Nat} A_n
$
for the decreasing sequence $\Folge{A}$ in $\mathcal{A}$. 
\EndProof

We  relate $Q\in\Vau{S}$ to the characteristic relation $R$ on $S$ by
comparing $\bas{A}{\geq q}\in Q$ with $\langle q, A\rangle\in R$ by
imposing a syntactic and a semantic condition. They will be shown to be equivalent. 

\BeginDefinition{satisf}
$Q\in\Vau{S}$ is said to \emph{satisfy} the characteristic relation $R$ on $S$ ($Q \vdash R$) iff we have 
\begin{equation*}
\langle q, A\rangle\in R \Leftrightarrow \bas{A}{\geq q}\in Q
\end{equation*}
for any $q\in [0, 1]$ and any $A\in\mathcal{A}$.
\EndDefinition

This is a syntactic notion. Its semantic counterpart reads like this:

\BeginDefinition{implem}
$Q$ is said to \emph{implement} $\mu\in\SubProb{S}$  iff 
\begin{equation*}
\mu(A) \geq q \Leftrightarrow \bas{A}{q}\in Q
\end{equation*}
for any $q\in [0, 1]$ and any $A\in\mathcal{A}$. We write this as $Q \models\mu$.
\EndDefinition

Note that $Q\models \mu$ and $Q \models \mu'$ implies 
\begin{equation*}
\forall A \in \mathcal{A}\forall q \geq 0: \mu(A) \geq q \Leftrightarrow \mu'(A) \geq q.
\end{equation*}
Consequently, $\mu = \mu'$, so that the measure implemented by $Q$ is uniquely determined. 

We will show now that syntactic and semantic issues are equivalent:
$Q$ satisfies a characteristic relation if and only if it implements
the corresponding measure. This will be used in a moment for a
characterization of those game frames which are generated from Kripke
frames.

\BeginProposition{satisf=implem}
$Q \vdash R$ iff $Q \models \mu_R$.
\EndProposition

\BeginProof
``$Q \vdash R~\Rightarrow~Q \models \mu_R$'':
Assume that $Q \vdash R$ holds. It is then immediate that $\mu_R(A) \geq r$ iff 
$\bas{A}{\geq r} \in Q$.

``$Q \models \mu_R~\Rightarrow~Q \vdash R$'': If $Q \models \mu_R$ for
relation $R \subseteq [0, 1] \times \mathcal{A}$, we establish that the
conditions given in Definition~\ref{R-binding} are
satisfied~\cite[Proposition 4.1.23]{EED-Companion}.
\begin{enumerate}
\item  Let $\bas{A}{\geq r}\in Q$ and $A \subseteq B$, thus $\mu_R(A) \geq r$, hence $\mu_R(B) \geq r$, which in turn implies $\bas{B}{\geq r}\in Q$. Hence $\text{\ding{172}}$ holds. $\text{\ding{173}}$ is established similarly.
\item If $\mu_R(A) < r$ and $\mu_R(B) < s$ with $r + s \leq 1$, then 
$
\mu_R(A\cup B) = \mu_R(A) + \mu(B) - \mu_R(A \cap B) \leq \mu_R(A) + \mu_R(B) < r + s,
$
which implies $\text{\ding{174}}$. 
\item If $\mu_R(A\cup B) \geq r$ and $\mu_R(A \cup (S\setminus B)) \geq s$, then 
$
\mu_R(A) = \mu_R(A\cup B) + \mu_R(A \cup (S\setminus B)) \geq r + s,
$
hence $\text{\ding{175}}$.
\item Assume $\mu_R(A) \geq r$ and $r + s > 1$, then 
$
\mu_R(S\setminus A) = \mu_R(S) - \mu_R(A) < p,
$
thus $\text{\ding{176}}$ holds. 
\item
If $\mu_R(\emptyset) \geq r$, then $r = 0$, yielding $\text{\ding{177}}$. 
\item 
Finally, if $\Folge{A}$ is decreasing with $\mu_R(A_n) \geq r$ for each $n\in\Nat$, then it is plain that
$
\mu_R(\bigcap_{n\in\Nat}A_n) \geq r.
$
This implies $\text{\ding{178}}$. 
\end{enumerate}
\EndProof

This permits a complete characterization of those stochastic effectivity functions which are generated through stochastic relations.

\BeginProposition{GtoK}
Let $P$ be a stochastic effectivity frame on state space $S$. Then these conditions are equivalent
\begin{enumerate}
\item\label{GtoK-1} There exists a stochastic relation $K: S \Trans S$ such that $P = P_K$.
\item\label{GtoK-2} $R(s) := \{\langle r, A\rangle \mid \bas{A}{\geq r}\in P(s)\}$ defines  a characteristic relation on $S$ with $P(s) \vdash R(s)$ for each state $s \in S$.
\end{enumerate}
\EndProposition

\BeginProof
``\labelImpl{GtoK-1}{GtoK-2}'':
Fix $s \in S$.  Because 
$
\bas{A}{\geq r}\in P_K(s)
$
iff
$
K(s)(A) \geq r,
$
we see that 
$
P(s) \models K(s),
$
hence by Proposition~\ref{satisf=implem}
$
P(s) \vdash R(s).
$

``\labelImpl{GtoK-2}{GtoK-1}'': Define $ K(s) := \mu_{R(s)}, $ for
$s \in S$, then $K(s)$ is a subprobability measure on
$\mathcal{A}$. We show that $K: S \Trans S$. Let
$G \subseteq \SubProb{S}$ be a *-measurable set, then
$ G \times [0, 1]\subseteq\SubProb{S}\times[0, 1] $ is measurable,
hence the measurability condition on $P$ yields that
\begin{equation*}
\InvBild{K}{G}  = \{s \in S \mid K(s)\in G\}
 = \{s \in S \mid G \in P(s)\}
\end{equation*}
is a measurable subset of $S$, because 
\begin{equation*}
\{\langle s, q\rangle \mid (G \times [0, 1])_q\in P(s)\} 
=
\{s \in S \mid G\in P(s)\} \times [0, 1]
\subseteq S\times[0, 1]\text{ is measurable}.
\end{equation*}
\EndProof
\subsection{Game Frames}

We define \emph{game frames} similar to Kripke frames  as being comprised of a
state space and the maps which indicate the actions to be taken. 

\BeginDefinition{G-frame}
A \emph{game frame} $\Ge = \bigl(S, (P_\gamma)_{\gamma\in\Gamma}\bigr)$ has a
measurable space $S$ of states and  a t-measurable map $P_\gamma: S
\to \Vau{S}$\MMP{Game frame} for each primitive game $\gamma\in\Gamma$.
\EndDefinition

Now that we have game frames, we can do some interesting things,
e.g., use them for the interpretation of game logics (well, nearly).
What we do first is to define recursively a set valued function
$\dasI{\tau}{A}{q}$\MMP{$\dasI{\tau}{A}{q}$} with the intention to describe the set of states
for which Angel upon playing game $\tau$ has a strategy of reaching a
state in set $A$ with probability greater than $q$. Assume that
$A\in\Borel{S}$ is a measurable subset of $S$, and $0 \leq q < 1$, and
define for $0 \leq k \leq \infty$
\begin{equation*}
Q^{(k)}(q) := \{\langle a_1, \dots, a_k\rangle \in \Rational^k \mid a_i \geq 0\text{ and } \sum_{i=1}^k a_i \leq q\}.
\end{equation*}
as the set of all non-negative rational $k$-tuples resp. sequences the
sum of which does not exceed $q$.
\begin{dingautolist}{202}
\item\label{primitive} Let $\gamma\in\Gamma$ be a primitive game, then\MMP{$\gamma\in\Gamma$} put
\begin{equation*}
\dasI{\gamma}{A}{q} := \{s\in S \mid \bas{A}{q}\in P_\gamma(s)\}, 
\end{equation*}
in particular
\begin{equation*}
\dasI{\epsilon}{A}{q} = \{s \in S \mid \delta_s(A) \geq q\} = A.
\end{equation*}
Thus
$s \in \dasI{\gamma}{A}{q}$ iff Angel has $\bas{A}{q}$ in its
portfolio when playing $\gamma$ in state $s$. This implies that the
set of all state distributions which evaluate at $A$ with a
probability greater than $q$ can be effected by Angel in this
situation. If Angel does not play at all, hence if the game $\gamma$
equals $\epsilon$, nothing is about to change, which means
\begin{equation*}
\dasI{\epsilon}{A}{q} = \{s \mid \delta_s\in\bas{A}{q}\} = A,
\end{equation*}
as expected. 
\item\label{dual} Let $\tau$ be a game, then\MMP{$\tau^d$} 
\begin{equation*}
\dasI{\tau^d}{A}{q} := S \setminus\dasI{\tau}{S \setminus A}{q}.
\end{equation*}
The game is determined, thus Demon can reach a set of states iff Angel does not have a strategy for reaching the complement. Consequently, upon playing $\tau$ in state $s$, Demon can reach a state in $A$ with probability greater than $q$ iff Angel cannot reach a state in $S\setminus A$ with probability greater $q$. 

Illustrating, let us assume for the moment that
$P_\gamma = P_{K_\gamma}$, i.e., the special case in which the
effectivity function for the primitive game $\gamma\in\Gamma$ is generated from a
stochastic relation $K_\gamma$. Then
\begin{equation*}
s \in \dasI{\gamma^d}{A}{q}
\Leftrightarrow
s \notin \dasI{\gamma}{S \setminus A}{q} 
\Leftrightarrow
K_\gamma(s)(S\setminus A) \leq q. 
\end{equation*}
In general, 
\begin{equation*}
s \in \dasI{\gamma^d}{A}{q}
\Leftrightarrow
\bas{S\setminus A}{q} \notin P_\gamma(s)
\end{equation*}
for $\gamma\in\Gamma$. This is exactly what one would expect in a determined game.
\item\label{angelic-choice}
Assume $s$ is a state such that Angel has a strategy for reaching a state in $A$ when playing the game $\tau_1\cup\tau_2$ with probability not greater than $q$. Then  Angel should have a strategy in $s$ for reaching a state in $A$ when playing game  $\tau_1$ with probability not greater than $a_1$ and playing game  $\tau_2$ with probability not greater than $a_2$ such that $a_1 + a_2 \leq q$. Thus\MMP{$\tau_1\cup\tau_2$}
\begin{equation*}
\dasI{\tau_1\cup\tau_2}{A}{q} := \bigcap_{a \in Q^{(2)}(q)}\bigl(\dasI{\tau_1}{A}{a_1}\cup\dasI{\tau_2}{A}{a_2}\bigr).
\end{equation*}
\item\label{Left-distr} 
Right distributivity of composition over angelic choice translates to this equation.\MMP{$(\tau_1\cup\tau_2);\tau$}
\begin{equation*}
\dasI{(\tau_1\cup\tau_2);\tau}{A}{q} := \dasI{\tau_1;\tau\cup\tau_2;\tau}{A}{q}. 
\end{equation*}
\item\label{Left-times} If $\gamma\in\Gamma$, put\MMP{$\gamma;\tau$}
\begin{equation*}
\dasI{\gamma;\tau}{A}{q} := \{s \in S \mid G_\tau(A, q) \in P_\gamma(s)\},
\end{equation*}
where
\begin{equation*}
G_\tau(A, q) := \{\mu\in\SubProb{S} \mid \int_0^1\mu(\dasI{\tau}{A}{r})\ dr > q\}.
\end{equation*}
Suppose that $\dasI{\tau}{A}{r}$ is already defined for each $r\in[0, 1]$ as the set of states for which Angel has a strategy to effect a state in $A$ through playing $\tau$ with probability greater than $r$. Given a distribution $\mu$ over the states, the integral 
$
\int_0^1\mu(\dasI{\tau}{A}{r})\ dr
$
is the expected value for entering a state in $A$ through playing $\tau$ for $\mu$. The set $G_\tau(A, q)$ collects all distributions the expected value of which is greater that $q$. We ask for all states such that Angel has this set in its portfolio when playing $\gamma$ in this state. Being able to select this set from the portfolio means that when playing $\gamma$ and subsequently $\tau$ a state in $A$ may be reached with probability greater than $q$.
\item\label{star}
The transformation for $\tau^{*};\tau_{0}$ is obtained through a
fairly direct translation of assumption~\ref{star-assumption}. by repeated application of the rule \ref{angelic-choice}. for angelic choice\MMP{$\tau^*;\tau_0$}:
\begin{equation*}
\dasI{\tau^*;\tau_0}{A}{q} := \bigcap_{a\in Q^{(\infty)}(q)}\bigcup_{n\geq 0}\dasI{\tau^n;\tau_0}{A}{a_{n+1}}
\end{equation*}
with $\tau^n := \tau;\dots;\tau$ ($n$ times).
\end{dingautolist}

We obtain for state spaces that are closed under the Souslin
operations (see Appendix~\ref{sec:souslin}).
\BeginProposition{prop-interp}
Assume that the state space is closed under the Souslin operation,
then  $ \dasI{\tau}{A}{q}$ is a measurable subset of $S$ for
all $A\subseteq S$ measurable, for all games $\tau$, and
$0 \leq q \leq 1$.
\QED
\EndProposition

Note that universally complete measurable spaces or analytic sets, which are
popular in some circles, are closed under this mysterious
operation (in fact, you can represent each analytic set in a Polish
space through a
Souslin scheme of closed sets~\cite[Proposition 4.5.6]{EED-Companion}), but other important spaces like Polish spaces are not. 

Finally, we introduce \emph{game models}; they are what you expect. Take a
game frame and add for each primitive formula the set of all states in
which it is assumed to be valid. 

\BeginDefinition{game-model}
A \emph{game model} $\Ge = \bigl(S, (P_\gamma)_{\gamma\in\Gamma}, (V_p)_{p\in \AE}\bigr)$
over measurable space $S$ is given by a game frame $ \bigl(S, (P_\gamma)_{\gamma\in\Gamma}\bigr)$, and by
a family $(V_p)_{p\in \AE}$ of sets which assigns to each atomic
statement a measurable set of state space $S$. We denote the
underlying game frame by $\Ge$\MMP{Game model} as well.
\EndDefinition
\renewcommand{\Gilt}[2][\phi]{\LinkeKlammer#1\RechteKlammer_{#2}}
Define the validity sets for each formula recursively as follows:
\begin{align*}
\Gilt[\top]{\Ge} & := S\\
\Gilt[p]{\Ge} & := V_p, \text{ if } p \in \AE\\
\Gilt[\phi_1\wedge\phi_2]{\Ge} & :=   \Gilt[\phi_1]{\Ge}\cap\Gilt[\phi_2]{\Ge}\\
\Gilt[\langle \tau \rangle_q \phi]{\Ge} & := \dasI{\Gilt{\Ge}}{\tau}{q}
\end{align*}

Accordingly, we say that formula $\phi$ holds in state $s$ ($\Ge, s \models \phi$) iff $s \in \Gilt{\Ge}$.

The definition of $\Gilt[\langle \tau \rangle_q \phi]{\Ge}$ has a
coalgebraic flavor. Coalgebraic logics define the validity of modal
formulas through special natural transformations (called
\emph{predicate liftings}) associated with the
modalities~\cite{Moss,Schroeder-Expressivity,EED-CoalgLogic-Book}. You
may wish to look up the brief discussion in~\cite[Section
2.7.3]{EED-Companion}. 
   
\BeginProposition{gilt-is-meas}
If state space $S$ is closed under the Souslin operation, $\Gilt{\Ge}$ is a
measurable subset for all formulas $\phi$. Moreover, 
$
\{\langle s, r\rangle \mid s \in \Gilt[\langle \tau \rangle_r \phi]{\Ge}\}
$
is a  measurable subset of $S\times[0, 1]$.
\EndProposition

This shows that the transformations we consider do not leave the realm
of measurability, provided the space is decent enough.

\BeginProof
The proof proceeds by induction on the formula $\phi$. If $\phi = p
\in \AE$ is an atomic proposition, then the assertion follows from $V_p
\in \Borel{S}$. The induction step uses Proposition~\ref{prop-interp}.
\EndProof

\section{What To Do Next}
\label{sec:what-i-could}

The next question to discuss would be the expressivity of game
models. This is technically somewhat involved, because the base
mechanism is not exactly light footed. I refer you
to~\cite{EED-GameLogic-TR}. There also the relationship to Kripke
models is investigated and completely characterized. The bridge to
stochastic nondeterminism through what is called \emph{hit
  measurability} from~\cite{Terraf-Bisim-MSCS} is constructed
in~\cite{EED+PST} with some new results on bisimilarity.

\begin{appendix}
\section{The $\pi$-$\lambda$-Theorem}
\label{sec:pi-lambda-theorem}
The measure theoretic reason why we insist on our modal logics being closed under conjunctions is Dynkin's
  famous \emph{$\pi$-$\lambda$-Theorem}: 
\BeginTheorem{Pi-Lambda}
Let $\mathcal{P}$ be a family of
subsets of of a set $S$ which is closed under finite intersections. Then $\sigma(\mathcal{P})$ is the smallest
class containing $\mathcal{P}$ which is closed under complements and countable disjoint unions.
\QED\EndTheorem

For a proof, see~\cite[Theorem 1.6.30]{EED-Companion}. There you find also a
discussion on its use, e.g., when establishing the equality of
measures from generators of $\sigma$-algebras to the $\sigma$-algebra proper.

The application to modal logics is immediate, given that conjunction
of formulas translates to the intersection of the validity sets.

\section{The Souslin Operation}
\label{sec:souslin}

\def\wrd#1{{#1}^*}
$\wrd{V}$ denotes for a set $V$ the set of all finite words with letters from $V$ including the empty string $\epsilon$. Let 
$
\{A_s\mid s\in \wrd{\Nat}\}
$
be a collection of subsets of a set $X$ indexed by all finite sequences of natural numbers (a \emph{Souslin scheme}), then the \emph{Souslin operation $\mathfrak{A}$} on this collection is defined as
\begin{equation*}
\mathfrak{A}\bigl(\{A_s\mid s\in \wrd{\Nat}\}\bigr) := \bigcup_{\alpha\in\Nat^{\Nat}}\bigcap_{n\in\Nat} A_{\alpha|n},
\end{equation*}
where $\alpha|n\in\wrd{\Nat}$ is just the word composed from the first $n$ letters of the sequence $\alpha$. This operation is intimately connected with the theory of analytic sets~\cite{Jech,Bogachev,Srivastava}. We obtain from~\cite[Proposition 1.10.5]{Bogachev}:

\BeginProposition{closed-under-Souslin}
If $S$ is a universally complete measurable or an analytic space, then its measurable
sets are closed under the operation $\mathfrak{A}$. 
\QED
\EndProposition

A discussion of the Souslin operation together with some technical
tools associated with it in given in~\cite[Section
4.5]{EED-Companion}.

\end{appendix}

\paragraph{Acknowledgements.}
The tutorial on which these notes are based was presented in December
2015 and January 2016 at the Department of Mathematics and Computer
Science of the University of Udine. I want to thank Prof. Marino
Miculan and Prof. Alberto Policritti for inviting me to present this
tutorial to their students; Prof. Eugenio Omodeo provided as always some helpful
comments, for which I want to thank him.  

\bibliographystyle{plain}

\end{document}